\documentclass[twocolumn,showpacs,preprintnumbers,amsmath,amssymb,aps,prb,superscriptaddress]{revtex4-1}

\pdfoutput=1
\usepackage[pdftex]{graphicx}
\usepackage[english]{babel}
\usepackage{soul}
\usepackage{cleveref}
\usepackage{amsmath,esint}
\usepackage{multirow}

\usepackage{dcolumn}
\usepackage{bm}

\usepackage{color}
\usepackage{amstext}
\usepackage{braket}

\begin{document}


\title{Topological bound states in interacting Su-Schrieffer-Heeger rings}

\author{A. M. Marques}
\email{anselmomagalhaes@ua.pt}
\affiliation{Department of Physics $\&$ I3N, University of Aveiro, 3810-193 Aveiro, Portugal}
\author{ R. G. Dias}
\affiliation{Department of Physics $\&$ I3N, University of Aveiro, 3810-193 Aveiro, Portugal}

\date{\today}


\begin{abstract}
We study two-particle states in a Su-Shrieffer-Heeger (SSH) chain with periodic boundary conditions and nearest-neighbor (NN) interactions.
The system is mapped into a problem of a single particle in a two-dimensional (2D) SSH lattice with potential walls along specific edges. 
The 2D SSH model has a trivial Chern number but a non-trivial Zak's phase, the one-dimensional (1D) topological invariant, along specific directions of the lattice, which allow for the presence of topological edge states. 
Using center-of-mass and relative coordinates, we calculate the energy spectrum of these two-body states for strong interactions and find that, aside from the expected appearance of doublon bands, two extra in-gap bands are present.
These are identified as bands of topological states localized at the edges of the internal coordinate, the relative distance between the two particles.
As such, the topological states reported here are intrinsically many-body in what concerns their real space manifestation, having no counterpart in single-particle states derived from effective models. 
Finally, we compare the effect of Hubbard interactions with that of NN interactions to show how the presence of the topological bound states is specific to the latter case.
\end{abstract}


\maketitle

\section{Introduction}
\label{sec:intro}

The interest in the topology of interacting systems has been gradually increasing in recent years. 
The topological characterization of its many-body states remains, to a large extent, an open problem, given that in the presence of interactions, topologically invariant quantities such as the Berry phase are, in general, ill-defined.
There is, however, a numerical approach which relies on the method developed by Niu and Thouless \cite{Niu1984}, where one imposes twisted boundary conditions on the many-body wavefunction of the ground state and then averages over all possible ones to recover a well defined Berry phase, $\gamma=n \pi$, where $n$ can be integer \cite{Guo2011,Grusdt2013,Li2015,Vanhala2016,Kumar2016} or fractional \cite{Guo2012,Budich2013}, depending on the filling factor and model considered. 

Two of the simpler ways to study many-body effects, in interacting 1D topological systems, consist of characterizing quasiparticle excitations at half-filling \cite{Kivelson1982,Guo2011,Liu2013,Grusdt2013,Weber2015} or two-body states \cite{Zhang2013,Bello2016,Liberto2016,Gorlach2017b}. 
Here we focus on the latter case.
It has been shown that interactions can drive the appearance of bands of bound states (doublons), in the energy spectrum as a function of the center-of-mass momentum $K$ \cite{Scott1994,Creffield2004,Valiente2008,Piil2008,Valiente2009,Javanainen2010,Wang2010,Creffield2010,Valiente2010,Nguenang2012,Longhi2013,Qin2014,Barbiero2015}.
Such bound states are expected to be experimentally accessible in optical lattices \cite{Wrinkler2006,Folling2007,Longhi2011,Krimer2011,Preiss2015,Mukherjee2016}. A distinct feature of these doublons is that their wavefunction is localized in the relative distance between its two component particles, which is immune to dissociation for strong interactions due to energy conservation.

In the context of SSH chains with open boundary conditions (OBC), the introduction of a strong Hubbard interaction produces midgap topological doublon edge states, upon application of a gate potential to compensate for a chemical potential shift at the edges \cite{Bello2016}.
Above a threshold value, this edge potential shift is responsible for an edge locking of bound states \cite{Pinto2009,Haque2010,Banchi2013}. 
The ability to either enhance or suppress this edge shift plays a crucial role in state transfer of bound states \cite{Compagno2017,Bello2017}.
In a previous paper \cite{Marques2017}, we considered instead the effects of NN interactions on two-hole excitations at half-filling.
We found some peculiar edge states, classified either as impurity-like, topologically originated (but not protected) or topologically protected in specific subspaces \cite{Xiao2017}.
Here, we extend this previous study to periodic boundary conditions (PBC), but focus on two-electron states (and not two-hole states as before).
We show that, in the strong-interaction regime, and besides the expected doublon energy bands, two extra doublon bands appear inside the gaps between the itinerant bands.
Their dependence on a finite difference in the value of the alternating hopping constants is a strong indicator of a decisive role being played by the non-trivial topology of the SSH model \cite{Su1979}.

Indeed, we find that the original problem of two spinless fermions with PBC can be reduced, through a series of \textit{exact} mappings using center-of-mass coordinates, to a one-particle problem in a topological chain with OBC for any $K$.
The open boundaries of the mapped chain arise in the new spatial coordinate considered, the relative distance between the two electrons.
For two specific momentum values, $K=0$ and $K=\pi$, we show how the extra bound states in the energy spectrum can be identified with the topological edge states of the mapped chains.
As we will address in more detail below, even though these extra bands of bound states have also appeared recently in Ref.~[\onlinecite{Liberto2017}], their topological origin is not discussed there, since they are characterized as analogs of a band of bound states present when on-site interactions are introduced. 
Here we show, on the other hand, that qualitative different features arise in the bound states as one changes from on-site to NN interactions.
More specifically, we explain why these topological bound states are not present with on-site interactions, being instead a distinctive feature of strong NN interactions.

Contrary to the bound edge states in open chains labeled topological in recent literature \cite{Bello2016,Bello2017,Gorlach2017b,Qin2017,Qin2018,Salerno2018}, where the bound state is described as a single particle in a topological state of an effective model of the open chain, the topological bound states in interacting SSH rings we report here have no counterpart in single-particle states and, thus, should be differentiated from those mentioned above. 
The specificity of the topological bound states found here comes from the fact that their localized behavior is not to be found at the edges of the original chain (as it is periodic). Instead, these states are localized in the \textit{internal} distance between the two particles, while extending spatially over the chain, making them intrinsically many-body states.

The rest of the paper is organized as follows. In Section~\ref{sec:model}, we introduce the model of an SSH ring with NN interactions and its respective mapping into a 2D SSH lattice.
In Section~\ref{sec:kpi}, we study the mapped lattice for $K=\pi$ and show the existence of topological states in the presence of strong NN interactions. 
In Section~\ref{sec:hubbard}, we drop the NN interactions and introduce instead a Hubbard interactions to show that topological states are absent in this latter case.
In Section~\ref{sec:k0}, we perform the same analyses as in the Section~\ref{sec:kpi} but now for $K=0$, and explicitly show the non-trivial topological nature of the model through the calculation of the Zak's phase.
Finally, in Section~\ref{sec:conclusions} we conclude.


\section{The model and 2D mapping}
\label{sec:model}

\begin{figure}[h]
\begin{center}
\includegraphics[width=0.42 \textwidth]{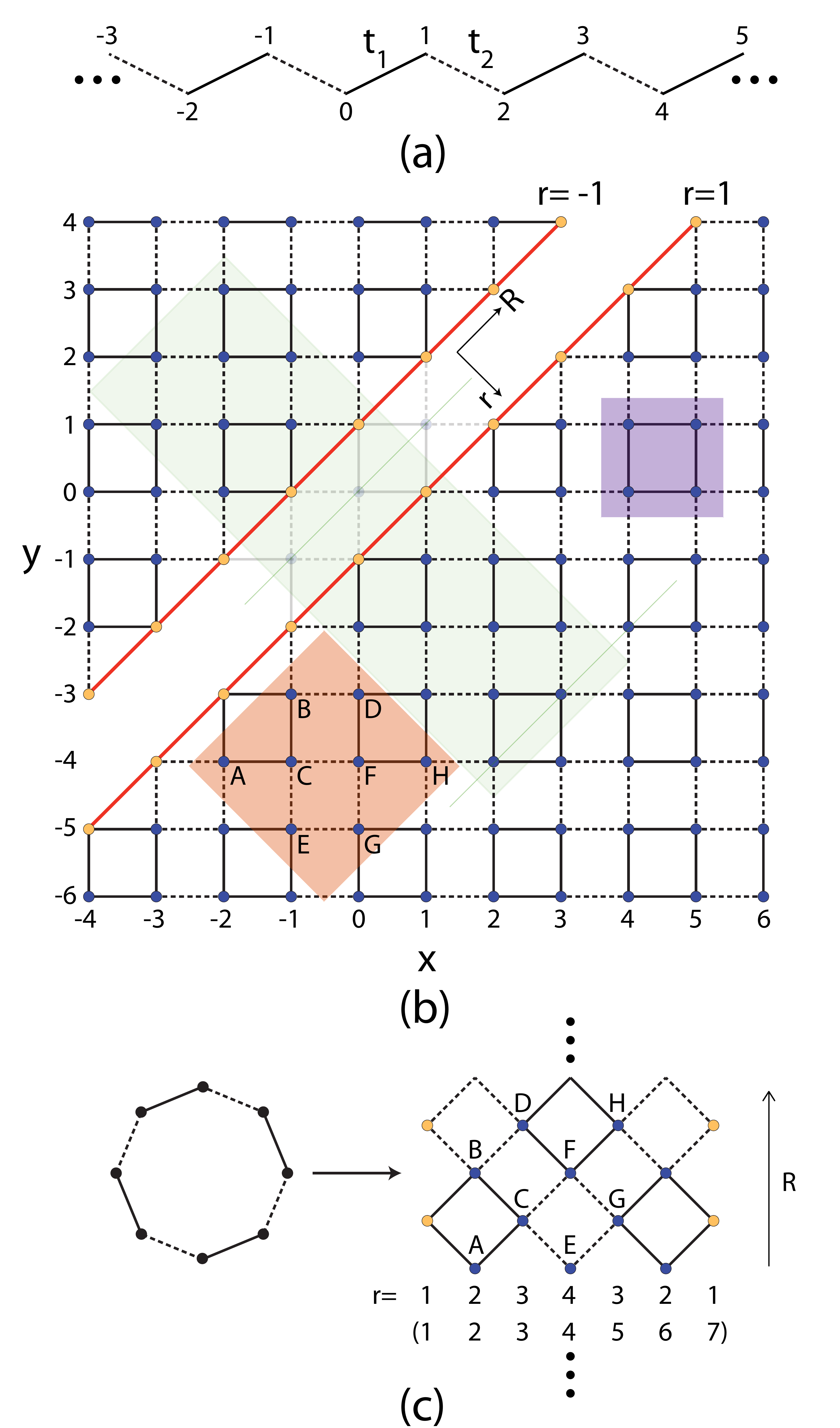}
\end{center}
\caption{(a) Infinite and periodic SSH model. 
(b) 2D mapping of the model of (a) for two spinless fermions and NN interactions, with one particle placed along each axis, $r$ being their internal distance and $R$ the center-of-mass. 
Sites along the $r=\pm 1$ lines have an on-site potential $V$ and the $r=0$ line is a forbidden region due to Pauli's exclusion principle. The purple square is the primitive cell and the red square with sites labeled A through H is the unit cell that repeats along the $r$ and $R$ directions.
The light green region indicates the mapped periodic strip of the $N=8$ SSH ring with two opposite spins and Hubbard interactions studied in Section~\ref{sec:k0}.
(c) Mapped strip of a finite periodic SSH chain with $N=2q=8$ sites. The complete mapped system consists of $q/2$ strips wrapped periodically in the $R$ direction. $r$ spans over $1\leq r\leq N-1=7$ (see Table~\ref{tab:statistics}), shown in parenthesis. In the SSH ring, one has $r\equiv N-r$, which is why it is convenient to span $r$ from 1 to $r_{max}=q$ and then back to 1, in order to keep the correspondence between problems.}
\label{fig:2dmapping}
\end{figure}

We consider a spinless SSH model of a periodic chain of $N$ sites with NN interactions, depicted in Fig.~\ref{fig:2dmapping}(a),
\begin{equation}
\label{eq:hamiltonian1d}
H=-\sum_{l=1}^N\big(t_l c^{\dagger}_{l}c_{l+1} +H.c.\big) 
+ V\sum_{l=1}^{N} n_{l}n_{l+1},
\end{equation}
where $l$ is the site number with $N+1\equiv 1$, $t_l=t_1$($t_2$) for $l$ odd (even) are the alternating hopping amplitudes, $n_{l}=c^{\dagger}_{l}c_{l}$ is the number operator, $V$ is the strength of the NN interaction and intercell spacing was set to $a=1$.
Throughout the paper, we set $t_1=1$ as the energy unit.
The problem of two interacting particles described by the 1D model of (\ref{eq:hamiltonian1d}) can be mapped onto a single-particle 
problem in a 2D lattice \cite{Longhi2011,Krimer2011,Corrielli2013,Mukherjee2016,Liberto2016,Gorlach2017,Gorlach2017b}
. 
In our case one arrives at the 2D SSH model of Fig.~\ref{fig:2dmapping}(b), with one of the particles of the original model placed along the $x$-axis and the other along the $y$-axis, so that a single particle in state $\ket{x,y}$ represents, in this 2D model, the global state of the two particles.
It is convenient to describe this system in new coordinates, using instead the center-of-mass $R=\frac{x+y}{2}$ and relative $r=x-y$ coordinates, together with the center-of-mass $K=k_x+k_y$ and relative $k=\frac{k_x-k_y}{2}$ momenta \cite{Valiente2008}. 
For two non-interacting particles in a finite SSH ring with $N$ sites, one can take advantage of certain symmetries of the states and of their periodicity [that is, $\ket{x,y}\equiv\ket{x\pm iN,y\pm jN}$ with $(i,j)\in \mathbb{Z}\times\mathbb{Z}$] to define a fundamental domain and its boundary conditions in the mapped 2D SSH lattice for each of the three distinct cases: 1) two distinguishable particles (e.g., two opposite spins), 2) two spinless fermions, and 3) two bosons.
The reader is referred to Appendix \ref{sec:appendix} for further details on the construction of the fundamental domain.
The results are condensed in Table~\ref{tab:statistics}, where one sees that the mapped lattice becomes a torus in case 1), with PBC in both the $R$ and $r$ directions, and a cylinder in cases 2) and 3), with PBC in the $R$ direction and OBC in the $r$ direction.
In 3), the bosonic case, an additional renormalization of the hoppings constants connecting $r=0$ and $r=1$ sites is required, as we will show in Section~\ref{sec:hubbard}.
\begin{table}[h]
\begin{center}
  \begin{tabular}{|c|c|c|}
\cline{3-3}
\multicolumn{2}{c|}{}&Distinguishable particles\\ 
\hline
\multirow{2}{*}{State}&in $xOy$&$\ket{x,y}=\ket{x}\otimes\ket{y}$
\\ 
\cline{2-3}
&in $rOR$&$\ket{R,r}=\ket{R+\frac{r}{2}}\otimes\ket{R-\frac{r}{2}}$\\
\cline{1-3}
\multirow{2}{*}{FD}&in $xOy$&$-x\leq y<-x+N\ \wedge\ x-N\leq y<x+N$\\
\cline{2-3}
&in $rOR$&$0\leq R<\frac{N}{2}\ \wedge\ -N<r\leq N$\\ 
\cline{1-3}
\multicolumn{2}{|c|}{Identifications}&$R\equiv R+\frac{N}{2}\ \wedge\ r\equiv r+2N$\\ 
\cline{1-3}
\multicolumn{2}{|c|}{Shape}&torus\\
\hline 
\multicolumn{3}{c}{}
\\
\cline{3-3}
\multicolumn{2}{c|}{}&Spinless fermions\\ 
\hline
\multirow{2}{*}{State}&in $xOy$&$\ket{x,y}_a=\frac{1}{\sqrt{2}}(\ket{x,y}-\ket{y,x})$
\\ 
\cline{2-3}
&in $rOR$&$\ket{R,r}_a=\frac{1}{\sqrt{2}}(\ket{R,r}-\ket{R,-r})$\\
\cline{1-3}
\multirow{2}{*}{FD}&in $xOy$&$-x-1\leq y<-x+N-1\wedge x-N+1\leq y\leq x-1$\\
\cline{2-3}
&in $rOR$&$-\frac{1}{2}\leq R<\frac{N-1}{2}\ \wedge\ 1\leq r\leq N-1$\\ 
\cline{1-3}
\multicolumn{2}{|c|}{Identifications}&$R\equiv R+\frac{N}{2}$\\ 
\cline{1-3}
\multicolumn{2}{|c|}{Shape}&cylinder\\
\hline 
\multicolumn{3}{c}{}
\\
\cline{3-3}
\multicolumn{2}{c|}{}&Bosons\\
\hline
\multirow{2}{*}{State}&in $xOy$&$\ket{x,y}_s=\frac{1}{\sqrt{2}}(\ket{x,y}+\ket{y,x})$
\\ 
\cline{2-3}
&in $rOR$&$\ket{R,r}_s=\frac{1}{\sqrt{2}}(\ket{R,r}+\ket{R,-r})$\\
\cline{1-3}
\multirow{2}{*}{FD}&in $xOy$&$-x-1\leq y<-x+N-1\wedge x-N\leq y\leq x$\\
\cline{2-3}
&in $rOR$&$-\frac{1}{2}\leq R<\frac{N-1}{2}\ \wedge\ 0\leq r\leq N$\\ 
\cline{1-3}
\multicolumn{2}{|c|}{Identifications}&$R\equiv R+\frac{N}{2}$\\ 
\cline{1-3}
\multicolumn{2}{|c|}{Shape}&cylinder\\
\hline 
  \end{tabular}  
\end{center}
\caption{Mapping into a 2D SSH lattice for three different two-particle systems of a non-interacting periodic SSH ring with $N$ sites.
FD is short for fundamental domain, whose method of construction is presented for the spinless fermions case in Appendix \ref{sec:appendix}. The identifications discriminate the periodic directions.}
\label{tab:statistics}
\end{table}

Focusing for now on the two spinless fermions case, same site occupation in 1D, that is, $r=0$ in 2D, is forbidden (an infinite on-site potential wall due to Pauli's exclusion principle).
Since we are considering identical fermions and the two half-planes in Fig.~\ref{fig:2dmapping}(b) are independent of each other, antisymmetric states with respect to the $r=0$ axis are equivalent, and so we use this to define the fundamental domain of our problem in the lower half-plane defined by $r\geq 1$ (see Table~\ref{tab:statistics}).
The NN interaction with strength $V$ in 1D is translated into a potential wall at the sites along the $r=\pm 1$ lines in the 2D SSH model.
In the extended SSH Bose-Hubbard model, a finite on-site Hubbard interaction $U$ introduces a finite potential wall at $r=0$. 
The interplay between these two types of walls, one for $r=0$ due to $U$ and the other for $r=\pm 1$ due to $V$, considerably enriches the two-particle energy spectrum \cite{Liberto2017}, particularly by the emergence of hybridized bound states.

\begin{figure}[h]
\begin{center}
\includegraphics[width=0.47 \textwidth, height=0.30 \textheight]{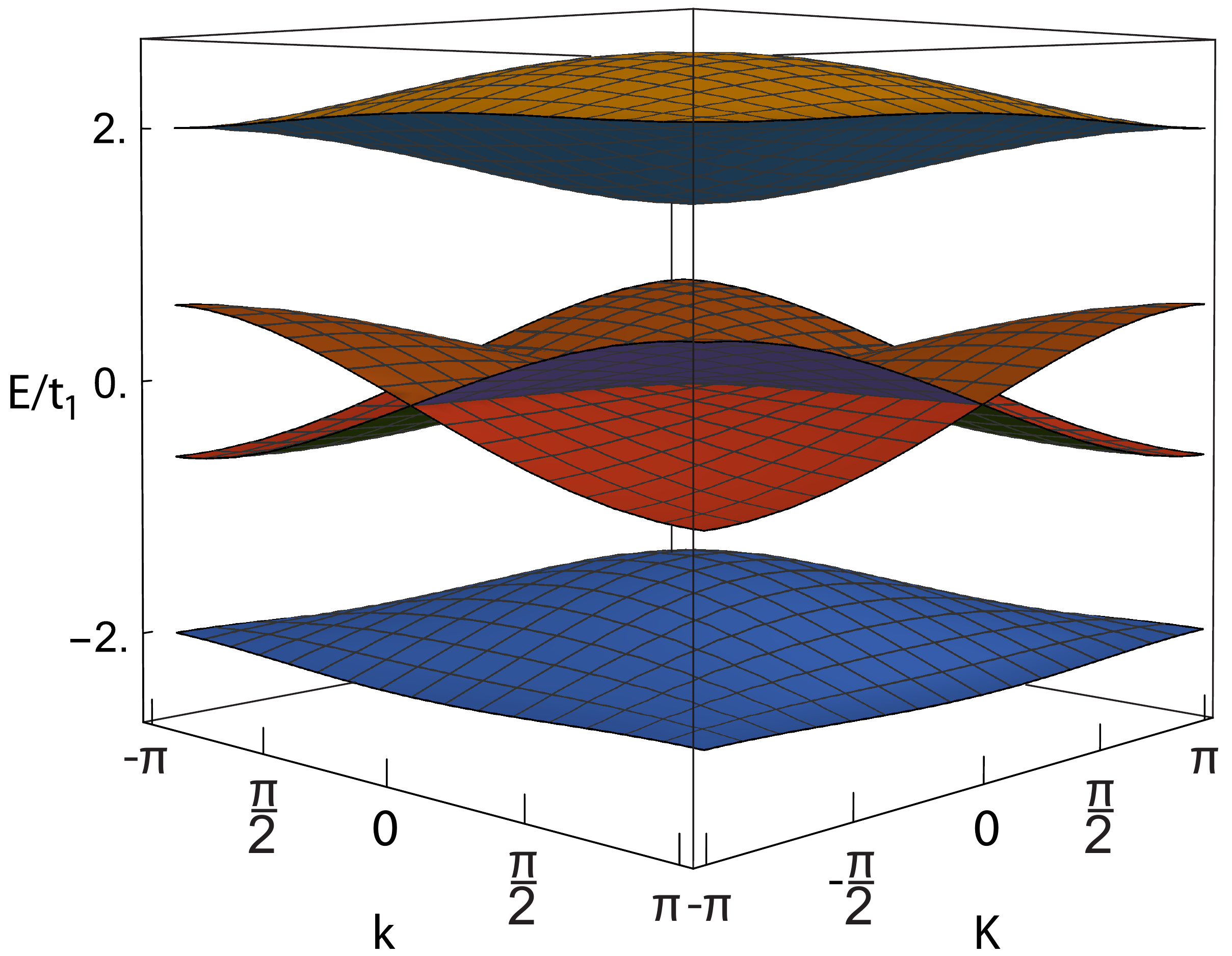}
\end{center}
\caption{Band structure of the 2D SSH model as a function of $K$ and $k$, the center-of-mass and relative momentum, respectively, for $t_2=0.3t_1$.
The system has a total of eight bands: two around $E/t_1=2$, two symmetric ones around $E/t_1=-2$ and two doubly degenerate bands around $E/t_1=0$, with one doubly degenerate band with $E/t_1\geq 0$ and another one with $E/t_1\leq 0$.}
\label{fig:2dsshedisperion}
\end{figure}

The red square with points labeled A through H in Fig.~\ref{fig:2dmapping}(b) corresponds to the unit cell of the mapped system that repeats in the $r$ and $R$ directions (the purple square corresponds to the primitive cell, repeating in the $x$ and $y$ directions).
This unit cell has double the energy bands (sites) of the primitive cell, which comes as a result of a folding of the Brillouin Zone of the primitive cell.
In the new momentum coordinates, the Hamiltonian of the 2D SSH chain, for this choice of unit cell, is given by
\begin{widetext}
\begin{equation}
H=
\begin{bmatrix}
0&0&t_1&t_2e^{-iK}&0&0&t_1e^{ik}&t_2e^{-i(K-k)}
\\
0&0&t_1&t_2&0&0&t_1e^{ik}&t_2e^{ik}
\\
t_1&t_1&0&0&t_2&t_2&0&0
\\
t_2e^{iK}&t_2&0&0&t_1e^{iK}&t_1&0&0
\\
0&0&t_2&t_1e^{-iK}&0&0&t_2&t_1e^{-iK}
\\
0&0&t_2&t_1&0&0&t_2&t_1
\\
t_1e^{-ik}&t_1e^{-ik}&0&0&t_2&t_2&0&0
\\
t_2e^{i(K-k)}&t_2e^{-ik}&0&0&t_1e^{iK}&t_1&0&0
\end{bmatrix}.
\end{equation}
\end{widetext}
Upon diagonalization of this Hamiltonian, one finds the energy spectrum of the system, as depicted in Fig.~\ref{fig:2dsshedisperion} for $t_2=0.3 t_1$.
Information about the topological nature of such 2D lattices is provided by the Chern number, defined for each band as
\begin{equation}
C=\frac{1}{2\pi}\oiint_{BZ} d\textbf{\textbf{k}}\mathcal{F}(\textbf{k}),
\end{equation}
where $\mathcal{F}(\textbf{k})=\nabla \times A(\textbf{k})$ is the Berry curvature, $A(\textbf{k})=- i \bra{\psi(\textbf{k})}\nabla_{\textbf{k}}\ket{\psi(\textbf{k})}$ is the Berry connection and $\ket{\psi(\textbf{k})}$ is the eigenstate of the band.

In our case, the fact that the 2D SSH model has time-reversal symmetry imposes that $\mathcal{F}(\textbf{k})=-\mathcal{F}(-\textbf{k})$ \cite{Xiao2010}.
Therefore, the integral over the Brillouin zone of this odd function vanishes, and so the Chern number is zero (note that the singularities of the Berry curvature at the degenerate energy points in Fig.~\ref{fig:2dsshedisperion}, which have to be treated separately, were also shown in Ref.[\onlinecite{Liu2017}] to integrate to zero over the Brillouin zone).
However, as has been shown in Ref.~[\onlinecite{Delplace2011}] for a system of weakly coupled SSH chains, a trivial Chern number does not necessarily imply a trivial topological insulator.
In this case, by turning on adiabatically the interchain coupling, which changes the system from a series of independent 1D chains to a 2D lattice, the topological nature of the isolated SSH chains is not changed, as it does not depend on the perpendicular direction of the interchain couplings (that is, if the SSH chains are in the topological phase, the existence of the correspondent topological edge states is not affected by turning on the interchain coupling).
As such, the 2D SSH model of Fig.~\ref{fig:2dmapping}(b) can be thought of as stacks of SSH chains with alternate interchain couplings in both the $x$ and $y$ directions.
The idea, then, is to carry over the information provided for each direction by the 1D topological invariant, given by Zak's phase, to the 2D lattice under consideration to determine the possible presence of topological states \cite{Hatsugai2009,Lau2015,Miert2017}.
This method can be straightforwardly generalized to systems of higher dimensionality \cite{Kariyado2013}.
Such 2D systems with trivial Chern number but non-trivial Zak's phase along certain 1D edges have also been referred to as weak 2D topological insulators \cite{Guo2014,Yoshimura2014,Matsumoto2015}
In the context of the 2D SSH model considered here, the presence of topological edge states has been recently linked to a non-trivial 2D Zak's phase \cite{Liu2017} (generalization of Zak's phase to 2D lattices).
The results we find below are consistent with this finding.

As shown in Table~\ref{tab:statistics}, a periodic 1D chain with two spinless fermions and finite size, when mapped onto a 2D plane with one particle, unfolds as a strip with open boundaries in the $r$ direction, wrapped periodically in the $R$ direction. 
This strip, the unit cell of this mapped system, is illustrated for the $N=8$ sites case in Fig.~\ref{fig:2dmapping}(c).
Notice that $N$ is always even, with $N=2q$ and $q=2,3,...$, since one can only add or remove pairs of $t_1$ and $t_2$ hoppings.
The complete 2D mapping for this case, shown at the left of Fig.~\ref{fig:fdtot1t1t2t2} in Appendix \ref{sec:appendix}, consists of a number of these strips, given by $\#_{strips}=q/2$, which is integer for $q$ even and half-integer for $q$ odd, forming a periodic cylinder shape around the $R$ direction (spanning over $N$ values for $R$, with $-\frac{1}{2}\leq R<\frac{N-1}{2}$ and $R=\frac{x+y}{2}$ varying in half-integer values).
One notices that the $V$ potential wall appears now at both ends of the strip. The internal distance between particles, along the $r$ direction, reaches a maximum, $r_{max}=q$, and then returns to one [see Fig.~\ref{fig:2dmapping}(c)].
The number of strips and the strip itself increase as $N$ increases, since $r_{max}=2\#_{strips}= q$ or, conversely, since the fundamental domain increases with $N$ in both the $R$ and $r$ directions (see Table~\ref{tab:statistics}).
If we considered a bosonic system with an interaction $U$ instead of $V$, the mapped lattice of Fig.~\ref{fig:2dmapping}(c) would become enlarged by the introduction of two extra sites with potential energy $U$ at $r=0$ on both ends of the strip.
As we will demonstrate below, this enlargement of the strip that occurs when one substitutes $V$ with $U$ plays a crucial role in preventing the appearance of topological bound states in the latter case.

\begin{figure}[h]
\begin{center}
\includegraphics[width=0.47 \textwidth]{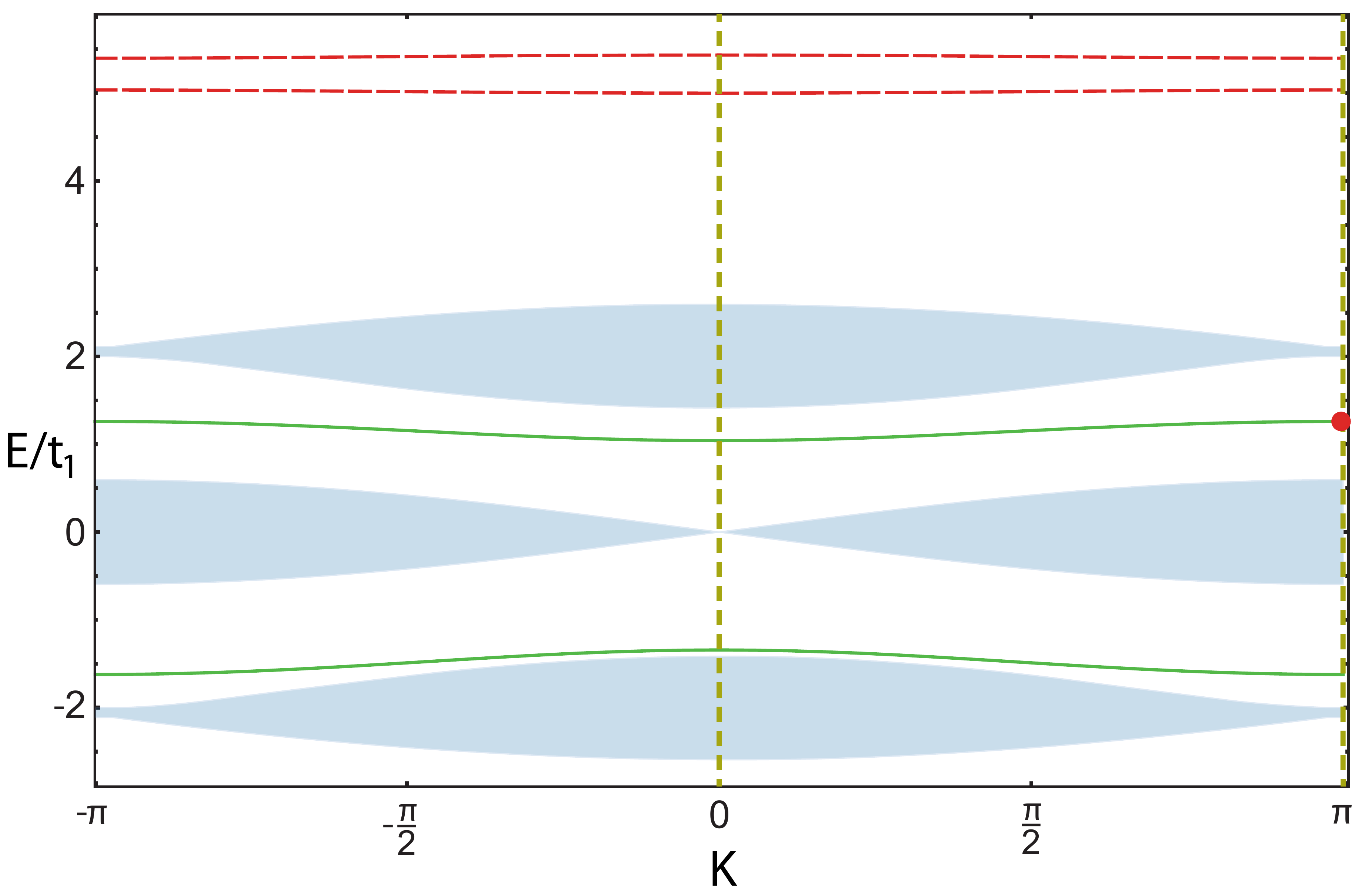}
\end{center}
\caption{Energy spectrum of a strip, for the case of two spinless fermions, as a function of the center-of-mass momentum in the continuum limit with parameters $(t_2,V)=t_1 (0.3,5)$.
Shaded blue regions correspond to the itinerant bands, given by a 2D projection of the bands of Fig.~\ref{fig:2dsshedisperion}, dashed red curves are doublon bands localized around $r=1$, that is, at the edge sites of the strip with on-site potential $V$, and solid green curves correspond to the bands of topological doublon edge states.
Vertical lines indicate the energy levels for the two inversion invariant center-of-mass momenta, at $K=0,\pi$.
The red dot at the upper topological band at $K=\pi$ indicates the state represented in Fig.~\ref{fig:unitcell}(c) for a finite strip with $N=10$.}
\label{fig:energydispersion}
\end{figure}

Since $K$ is a good quantum number along the periodic $R$ direction, one can Fourier transform the Hamiltonian of the mapped system in this direction to find its energy dispersion for this case of spinless fermions, shown in Fig.~\ref{fig:energydispersion}.
Aside from the expected doublon bands, associated with states localized around the $r=1$ sites with on-site potential $V$ (dashed red curves), one sees two additional bands of localized states (solid green curves) between the bands of itinerant states (shaded blue bands, given by a 2D projection of the bands of Fig.~\ref{fig:2dsshedisperion}). 
The solid green bands of Fig.~\ref{fig:energydispersion} correspond to the bands of $\tilde{d}_1$ and $\tilde{d}_2$ states in Fig.~2(b) of Ref.~[\onlinecite{Liberto2017}], for the set of parameters used there.

\begin{figure}[h]
\begin{center}
\includegraphics[width=0.47 \textwidth]{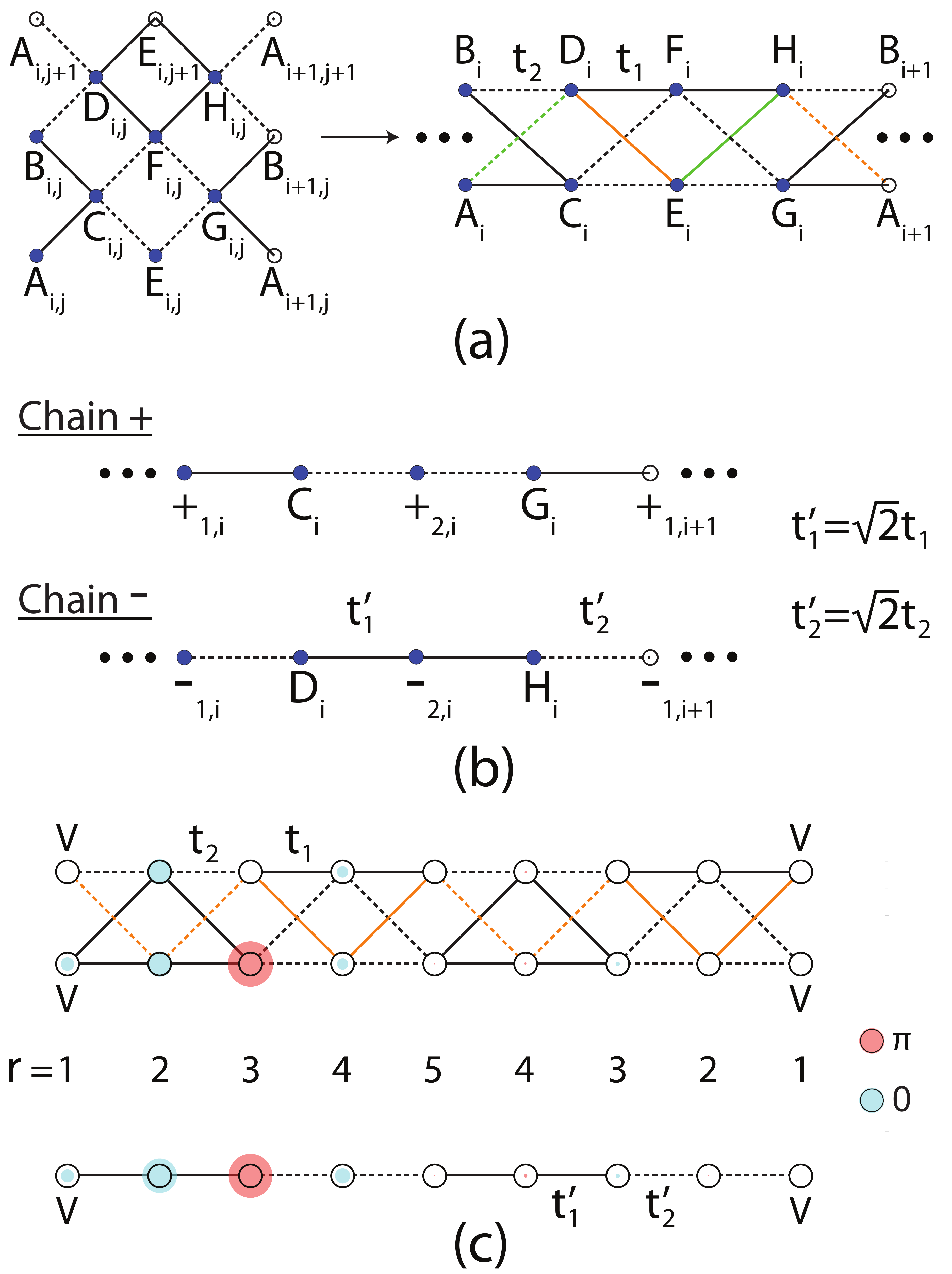}
\end{center}
\caption{(a) By Fourier transforming the Hamiltonian associated with unit cell of Fig.~\ref{fig:2dmapping}(b) along the $R$ direction, one obtains a 1D two-leg ladder with $K$-dependent phases across the colored hoppings, where the green and orange hoppings have an extra phase factor of $e^{-iK}$ and $e^{i K}$, respectively.
(b) Through the basis rotation given in (\ref{eq:basisrot1}) and (\ref{eq:basisrot2}), the two-leg ladder is decomposed, for $K=\pi$, in two decoupled chains with renormalized hopping constants.
(c) Spatial distribution of the topological bound state in the mapped two-leg ladder and ``chain +'' of the original periodic SSH chain with $N=10$ sites and $K=\pi$ for the case of spinless fermions, with energy given by the red dot in Fig.~\ref{fig:energydispersion} and for the same parameters considered there.
The orange hoppings have a minus sign (corresponding to a $\pi$ phase across these hoppings).
In each site, radius size and color indicate the relative amplitude of occupation and phase, respectively.
A degenerate topological state is present at the right end of the two-leg ladder, that is, at the right end of ``chain -''.}
\label{fig:unitcell}
\end{figure}

If we isolate the 1D Hamiltonian of the strip [which is an edge Hamiltonian obtained by cutting the bulk Hamiltonian of the 2D SSH lattice along the $r$ direction \cite{Ryu2002}, with the unit cell considered in Fig.~\ref{fig:2dmapping}(b)], we can calculate the Zak's phase for each $K$ value from the bulk properties of the strip as
\begin{equation}
\gamma(K)=-\sum_{n\in occ}i\int_{-\pi}^\pi dk \bra{u_n(k,K)}\frac{d}{dk}\ket{u_n({k,K})},
\label{eq:zaksphase}
\end{equation}
where $\ket{u_n({k,K})}$ is the eigenstate of band $n$, ``$occ$'' defines the set of occupied bands and $\gamma(K)$ is defined up to mod $2\pi$.
To highlight the topological nature of the in-gap states of Fig.~\ref{fig:energydispersion}, we will follow the following strategy: i) first, we Fourier transform the Hamiltonian of the strip in the $R$ direction to get the 1D chain with $K$-dependent hoppings given in Fig.~\ref{fig:unitcell}(a), ii) second, we consider the two inversion invariant center-of-mass momenta, $K=0,\pi$, for which Zak's phase is $\pi$-quantized \cite{Asboth2016}, and show how, for each of these momenta, the strip maps into a 1D topological system that harbors topological edge states for the set of $(t_2,V)$ values considered.

\section{$K=\pi$}
\label{sec:kpi}

For $K=\pi$ in the spinless fermions case, the $K$-dependence present in some of the inter-leg hoppings translates into a minus sign (a $\pi$ flux across these hoppings).
The basis of the two-leg ladder of Fig.~\ref{fig:unitcell}(a), under PBC, is given by $(\{\ket{\alpha_i}\})$, where $\alpha$ runs from A to H and $i$ is the unit cell index.
An appropriate rotation in this basis, given by the following symmetric and anti-symmetric combinations,
\begin{eqnarray}
\ket{\pm_{1,i}}&=&\frac{1}{\sqrt{2}}\big(\ket{B_i}\pm \ket{A_i}\big),
\label{eq:basisrot1}
\\
\ket{\pm_{2,i}}&=&\frac{1}{\sqrt{2}}\big(\ket{F_i}\pm \ket{E_i}\big),
\label{eq:basisrot2}
\end{eqnarray}
transforms the two-leg ladder into a system of two decoupled linear chains labeled ``chain +'' and ``chain -'', shown in Fig.~\ref{fig:unitcell}(b).
These linear chains correspond to two different choices for the unit cell in the $t_1t_1t_2t_2$ model that we have recently studied \cite{Marques2017b}.
We showed there how, due to the fact that the two possible inversion axes in each unit cell are never at its center [see Fig.~\ref{fig:unitcell}(b)], a correction term has to be added to (\ref{eq:zaksphase}) in order to recover a $\pi$-quantized Zak's phase\cite{Kariyado2013}.
The general expression for the Zak's phase has the form
\begin{eqnarray}
\tilde{\gamma}(K)=
\begin{cases}
\begin{split}
\gamma(K) + \sum\limits_{n\in occ}\sum\limits_{s=1}^m\int_0^\pi dk \vert u_{n,s}(k,K)\vert^2, 
\\
\text{for\ } m>0,
\end{split}
\\
\\
\begin{split}
\gamma(K) - \sum\limits_{n\in occ}\sum\limits_{s=0}^{\vert m\vert -1}\int_0^\pi dk \vert u_{n,M-s}(k,K)\vert^2, 
\\
\text{for\ } m<0,
\end{split}
\end{cases}
\label{eq:generalmodzak}
\end{eqnarray}
where $m=0,\pm 1,...,\pm M$, with $M$ being the number of sites in the unit cell, indicates the displacement of the inversion axis, given by $r_m=a(\frac{1}{2}-\frac{m}{2M})$, from the center of the unit cell, while  $u_{n,j}(k,K)$, with $j=1,2,...,M$, represents the $j$-component of $\ket{u_n(k,K)}$.
Using (\ref{eq:generalmodzak}), the $t_1t_1t_2t_2$ model was shown be a topological insulator in Ref. [\onlinecite{Marques2017b}].

When $t_1>t_2$, as we assume here, in an open $t_1t_1t_2t_2$ chain, topological states are present at the ends where the end hopping is $t_1$, followed by a $t_2$ hopping [for example, in ``chain +'' of Fig.~\ref{fig:unitcell}(b), when the left end coincides with a $+_1$ site].
This should be contrasted, for example, with the SSH chain, where weak edge hoppings (which we also label $t_2$) are the condition for the presence of topological edge states.
In the limit where $t_2\to 0$, we get zero energy states localized at the isolated sites at the ends of the SSH model and, for the $t_1t_1t_2t_2$ model, the topological edge modes are bonding and anti-bonding states, with energies $\pm t_1$, of the two isolated sites at the edge connected by a $t_1$ hopping term.

However, one should remember that our original open system, aside from having the bulk form described by the two-leg ladder of Fig.~\ref{fig:unitcell}(a), is also terminated on both ends by two sites with on-site potential $V$ [see Fig.~\ref{fig:2dmapping}(c)], with each of them subsequently becoming the edge site of either ``chain +'' or ``chain -''.
The full system, with these edge sites included, has the general form of Fig.~\ref{fig:unitcell}(c), where the two-leg ladder and the ``chain +'' are depicted for the case of $N=10$, along with the spatial distribution of the topological state of higher energy (red dot in Fig.~\ref{fig:energydispersion}) correspondent to ``chain +'' (there is a degenerate state at the right edge of ``chain -'').

In the SSH model, an edge potential was shown to reverse the dimerization above a critical strength \cite{Rossi1992,Wada1992,Liberto2016,Marques2017}, by effectively separating the edge site from the rest of the chain.
In our $t_1t_1t_2t_2$ model, the effect of the edge potential is similar. 
For example, at the left edge of the ``chain +'' in Fig.~\ref{fig:unitcell}(c), the strong potential ($V=5t_1$) isolates the edge site at r=1, and the inner chain effectively starts at $r=2$, that is, the edge hopping of this inner chain is $t_1$ followed by a $t_2$ hopping, which is the condition for the presence of topological states for $t_1>t_2$, as mentioned above.
Furthermore, without the NN interaction $V$, the topological states would not be present for any periodic chain of size $N$.
This becomes clear if one considers that: 1) for any $N$, both ``chain +'' and ``chain -'' always have pairs of hoppings of the same type at each end, as exemplified in the ``chain +'' for $N=10$ in Fig.~\ref{fig:unitcell}(c); 2) since $N$ is even, to increase or decrease $N$ always involves the creation or destruction of pairs of hoppings of the same type at the ends of these chains; and 3) a sufficiently strong edge potential $V$ is therefore required to effectively isolate a $t_1$ hopping at specific edges of the inner chains, thus allowing for the presence of topological states, corresponding in Fig.~\ref{fig:energydispersion} to the in-gap states at the solid green curve for $K=\pi$.
Note that a chiral symmetry, defined as $CHC=-H$, where $C$ is assumed to be a local operator \cite{Ryu2002}, is recovered in the inner chain for $V\to \infty$. In this limit, the edge sites of the chain with potential energy $V$ become decoupled and can thus be projected out exactly from the inner chain Hamiltonian. 

However, $C$-symmetry is not, by itself, the protecting symmetry of the topological states (edge states are chiral pairs, their energies can change symmetrically when $C$-symmetry preserving perturbations are included). 
The definition of such a topologically protecting symmetry is somewhat trickier in this case. It involves the definition of a ``chiral-like'' operator that considers separately each pair of bands from which each edge state independently emerges. 
This new operator will turn out to be the chiral operator of the topological $H^2$ model (where $H^2$ represents the squared Hamiltonian of the $t_1t_1t_2t_2$ model, see an example of an $H^2$ model at Ref.~[\onlinecite{Kremer2018}]). 
A straightforward calculation shows that squaring the Hamiltonian of the $t_1t_1t_2t_2$ model yields two decoupled chains: an ionic Hubbard chain \cite{Valiente2010} and an SSH chain with a constant energy shift (given by the square of the energy of the edge states in the $t_1t_1t_2t_2$ model). 
There is a mapping between the topological edge states in the original $t_1t_1t_2t_2$ model and the topological edge states of the SSH chain of the squared model \footnote{A more in-depth discussion of these results will appear soon in a revised version of Ref.~[\onlinecite{Marques2017b}]}.

When switching from the unit cell with two consecutive $t_1$ hoppings to the one with a single $t_1$ hopping, the Zak's phase at each gap [sum of the Zak's phases of the bands below the gap, using (\ref{eq:generalmodzak})] of the $t_1t_1t_2t_2$ model is shifted by $\pi$, signaling the topological transition discussed above. Note that the Zak's phases of the two different unit cells can only be compared when one follows the same criterion for choosing the inversion axis, for example, choosing in both cases the inversion axis \textit{closer} to the center of the unit cell [with the smallest $|m|$ in (\ref{eq:generalmodzak})].

\section{Hubbard interaction for $K=\pi$}
\label{sec:hubbard}

We consider now the effects of dropping the $V$ term and introducing instead a Hubbard interaction $U$ in the Hamiltonian of (\ref{eq:hamiltonian1d}), which translates as an on-site potential at the sites across the $r=0$ diagonal in Fig.~\ref{fig:2dmapping}(b).
There are two distinct cases that have to be considered separately: i) the case of two distinguishable particles (we consider two opposite spins), and ii) the case of two identical bosons.
The Hubbard term that substitutes the NN interaction term in the Hamiltonian of (\ref{eq:hamiltonian1d}) is $H_U=U\sum_l n_{l\uparrow}n_{l\downarrow}$ for the former case (with the corresponding introduction of a spin index in the hopping term) and $H_U=\frac{U}{2}\sum_ln_l(n_l-1)$ for the latter case.

\begin{figure}[h]
\begin{center}
\includegraphics[width=0.47 \textwidth]{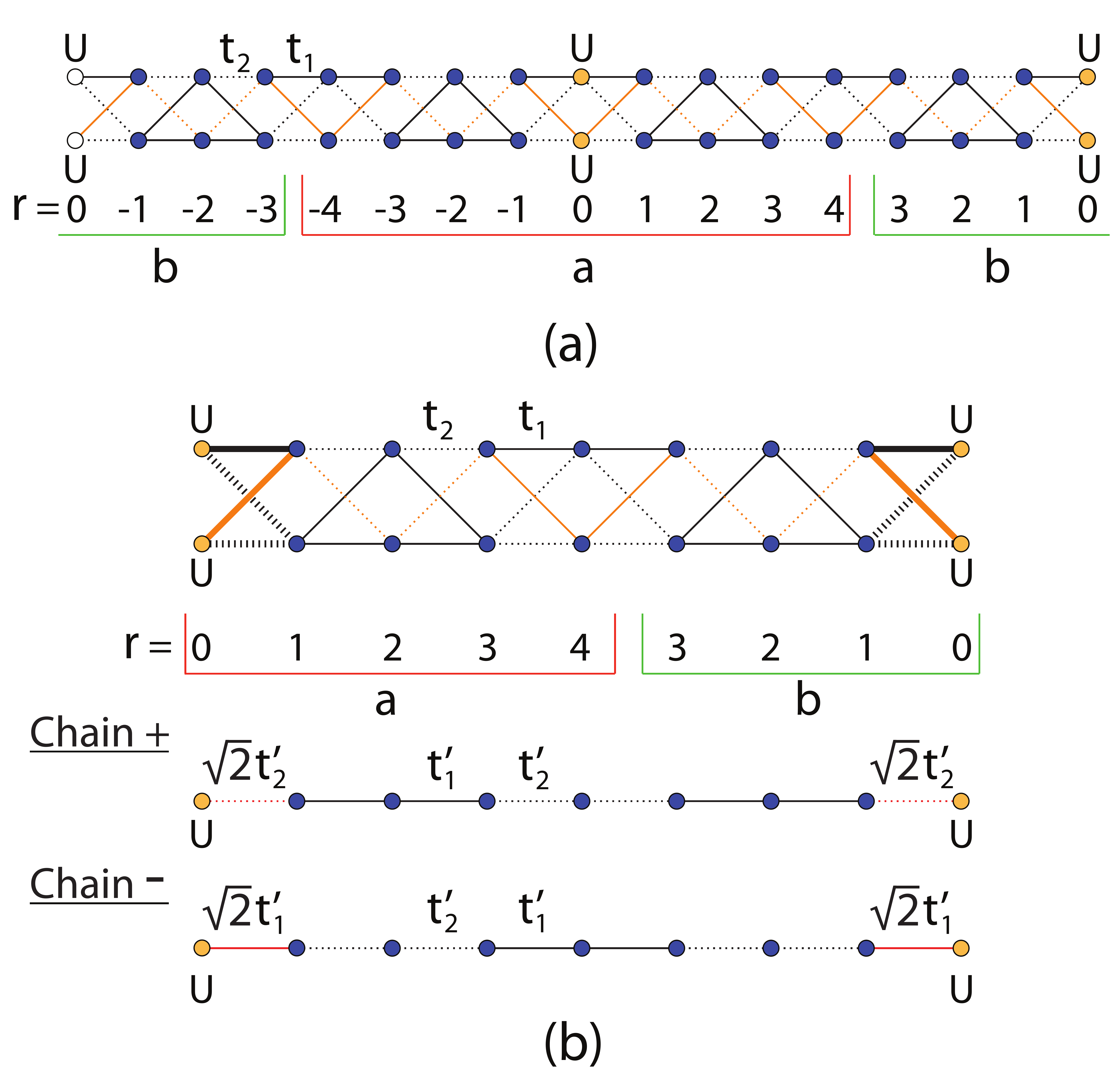}
\end{center}
\caption{Mapping of a finite periodic SSH chain with $N=8$, Hubbard interaction $U$ and $K=\pi$ for: (a) states of two opposite spins, which becomes a strip with PBC, and top of (b) two-boson states, which becomes a strip with OBC, where the thicker edge hopping constants have an extra factor of $\sqrt{2}$. 
Orange hopping constants have a minus sign.
We distinguish between regions $a$ and $b$ to facilitate site indexation.
Under the basis rotation given by (\ref{eq:basisrot1}) and (\ref{eq:basisrot2}), the strip in (b) separates into ``chain +'' and ``chain -'', with $t^\prime_i=\sqrt{2}t_i$}
\label{fig:hubbard}
\end{figure}

When we consider a system of two opposite spins, the full 2D mapping produces a lattice periodic in both the $r$ and $R$ directions with the shape of a torus (see Table~\ref{tab:statistics}).
Instead of an open strip in the $r$ direction, as in Fig.~\ref{fig:2dmapping}(c), one has now a periodic strip, as shown for the $N=8$ case in the light green region of Fig.~\ref{fig:2dmapping}(b), where the ends of this region are connected.
Upon Fourier transforming the respective Tight-binding Hamiltonian of this strip in the $R$ direction, as before, one arrives at the Hamiltonian of the two-leg ladder with PBC of Fig.~\ref{fig:hubbard}(a) for $K=\pi$.
The effect of large interactions, $U\gg t_i$, is to cut this ladder along both $r=0$ lines to produce an effective system of four decoupled sites with energy of the order of $U$ and two equivalent two-leg ladders with OBC at both ends.
An equivalent basis rotation to that of (\ref{eq:basisrot1}) and (\ref{eq:basisrot2}) transforms again each of these decoupled two-leg ladders into the ``chain +'' and ``chain -'' of Fig.~\ref{fig:hubbard}(b) without the yellow end sites (the inner chains with the blue sites only).
Given that both ``chain +'' and ``chain -'' always have pairs of consecutive hoppings of the same type at each end, there can be no topological bound states in this case, for the reasons explained above.
For these topological bound states to appear, one would have to introduce NN interactions, \textit{i.e.}, an on-site potential $V\gg t_i$ at the $r=1$ sites. Under these circumstances, and given that the inner chain starting from $r=2$ would be effectively decoupled from the two edge sites with on-site potentials $U$ and $V$, the value of $U$ has no effect on the topological states.

Let us consider now a system of two identical bosons for $K=\pi$ and $q=N/2$ even.
For a general $K$, it is convenient to write the states of the two-leg ladder of Fig.~\ref{fig:hubbard}(a) as 
\begin{equation}
\ket{K,r_u,\beta}=\frac{1}{\sqrt{N_R'}}\sum\limits_{R'} e^{-iKR'}\ket{R',r_u,\beta},
\end{equation}
where $u=a,b$ denotes the correspondent $r$ region in Fig.~\ref{fig:hubbard}(a), $R'=1,2,...,N_{R'}$, $N_{R'}=N/4$ is the number of different $R'$ values [each unit cell spans over 4 values for $R$, as can be seen, for example, in the A-C-B-D sequence in Fig.~\ref{fig:unitcell}(a)] and $\beta=t(b)$ gives the corresponding site at the top (bottom) leg of the ladder.
The symmetrized bosonic states have the form
\begin{equation}
\ket{K,r_u,\beta}_s=\frac{1}{\sqrt{2}}(\ket{K,r_u,\beta}+\ket{K,-r_u,\beta}),\ \ r_u>0.
\end{equation}
In the basis of these bosonic states, the system becomes once more an open two-leg ladder, shown at the top of Fig.~\ref{fig:hubbard}(b) for $K=\pi$.
Notice that the edge hoppings are renormalized, with an extra $\sqrt{2}$ factor, reflecting the fact that states $\ket{K,0_u,\beta}$ are already symmetrized.
For the $r_u=0$ sites at the top leg of Fig.~\ref{fig:hubbard}(a), the Tight-binding equation is written as
\begin{eqnarray}
H\ket{\pi,0_u,t}_s&=&t_1(\ket{\pi,1_u,t}+\ket{\pi,-1_u,t}) \nonumber
\\
&+&t_2(\ket{\pi,1_u,b}+\ket{\pi,-1_u,b}) \nonumber
\\
&=&\sqrt{2}t_1\ket{\pi,1_u,t}_s+\sqrt{2}t_2\ket{\pi,1_u,b}_s, 
\label{eq:symbosons}
\end{eqnarray}
with equivalent relations holding for the sites of the bottom leg at $r_u=0$.

Performing the basis rotation given by (\ref{eq:basisrot1}) and (\ref{eq:basisrot2}) for this case of two identical bosons, one arrives at the ``chain +'' and ``chain -'' of Fig.~\ref{fig:hubbard}(b) with $\sqrt{2}t_i$ hoppings at the edges.
The reasons which justify the absence of topological bound states in this bosonic case are not the same for the $U\to\infty$ and $U=0$ regimes.
Similarly to the opposite spins case, when $U\to\infty$ the edge sites of ``chain +'' and ``chain -'' become decoupled and the pairs of consecutive hoppings of the same type at the edges of the the inner chains prevent the appearance of the topological states.

When $U=0$, the problem becomes one of two non-interacting bosons.
Let us consider ``chain -'' for the case of $N=8$ shown in Fig.~\ref{fig:hubbard}(b), which has isolated $\sqrt{2} t^\prime_1$ hopping constants at the edges. 
We have seen that the condition for the existence of topological bound states is to have isolated $t^\prime_1$ hopping constants at one or both edges, for $t^\prime_1>t^\prime_2$ ($t_1>t_2$), as we have assumed throughout the paper.
However, the effect of the $\sqrt{2}$ factor at the edges has to be taken into account, as it is responsible for driving the energy of the topological states out of the gaps and into the energy bands, where they vanish. 
This can be better understood if one considers again the fully dimerized limit of $t^\prime_1>0$ and $t^\prime_2=0$ on the ``chain -'' of Fig.~\ref{fig:hubbard}(b).
The Hamiltonian describing the two sites at each edge is $H_{edge}=\sqrt{2}t^\prime_1\sigma_x$, where $\sigma_x$ is the first Pauli matrix, with eigenenergies $E_{edge,\pm}=\pm\sqrt{2}t^\prime_1$.
In the bulk, we have decoupled sites (at $r_u=2$) with $E=0$ and three sites connected by a pair of $t^\prime_1$ hopping constants (at $r_u=3$ and $r_a=4$).
In the basis spanned by the states $\{\ket{r_a=3},\ket{r_a=4},\ket{r_b=3}\}$, this bulk Hamiltonian is written as
\begin{equation}
H_{bulk}=
\begin{bmatrix}
0&t^\prime_1&0
\\
t^\prime_1&0&t^\prime_1
\\
0&t^\prime_1&0
\end{bmatrix},
\end{equation}
with eigenenergies $E_{bulk,0}=0$ and $E_{bulk,\pm}=\pm\sqrt{2}t^\prime_1$.
One sees that the extra $\sqrt{2}$ at the edge hoppings is precisely the only factor for which edge and bulk energies coincide, \textit{i.e.}, $E_{edge,\pm}=E_{bulk,\pm}$, and so the edge states vanish as the bulk bands broaden with the introduction of a finite $t^\prime_2$.
Note that this absence of topological bound states for $U=0$ comes purely as a consequence of Bose-Einstein statistics.
One way to circumvent this limitation would be to introduce $r$-modulated hopping constants, $t_1(r)=t_{1,r\leftrightarrow r+1}$, with $r\geq 0$, such that $t_1(0)\neq t_1(r)=t_1$ for $r>	0$.
Although it is not clear how this modulation could be experimentally realized, its implementation should in principle drive the appearance of topological bound states in this non-interacting regime.

\section{$K=0$}
\label{sec:k0}

\begin{figure}[h]
\begin{center}
\includegraphics[width=0.47 \textwidth]{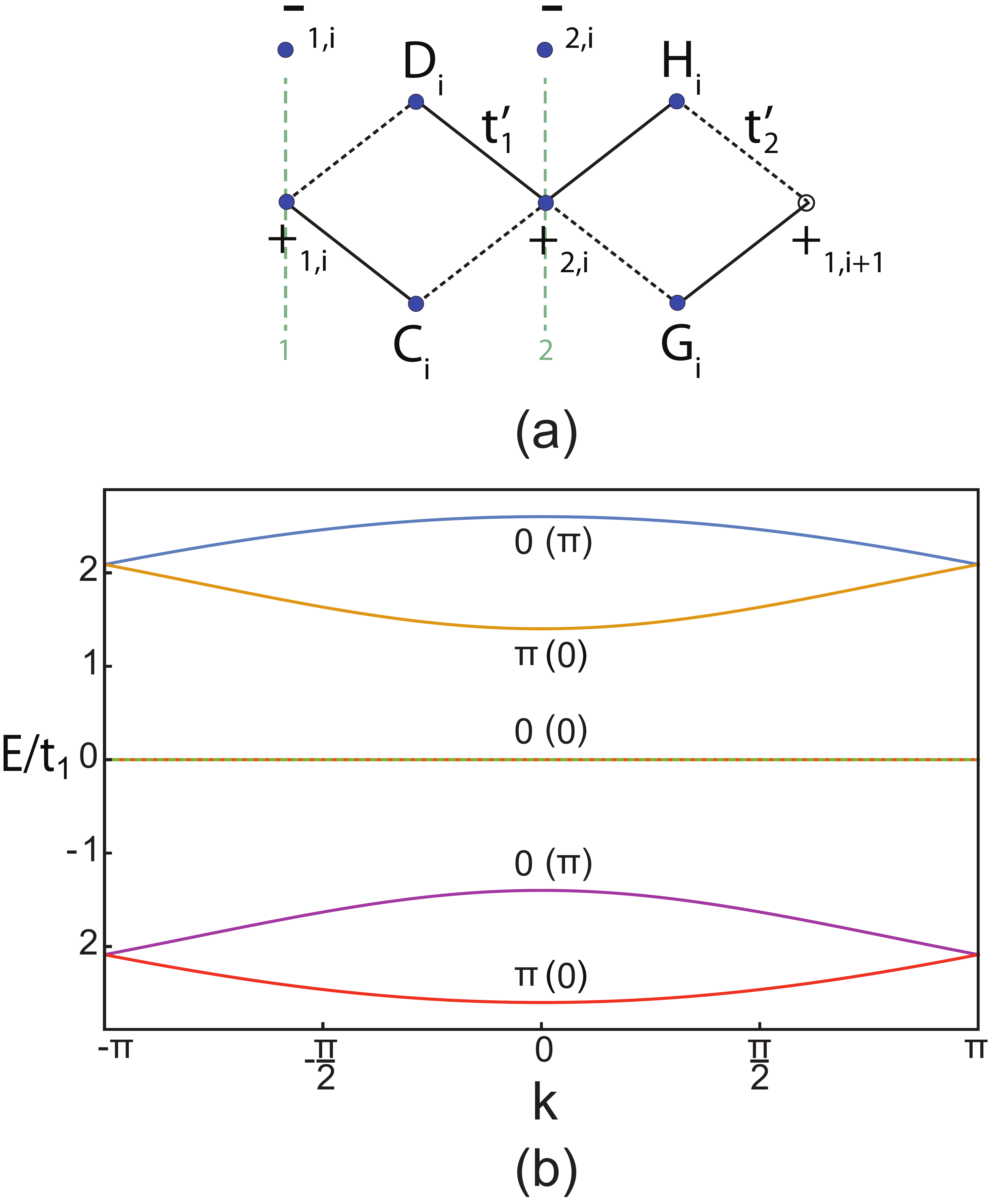}
\end{center}
\caption{(a) Through the basis rotation given in (\ref{eq:basisrot1}) and (\ref{eq:basisrot2}), the two-leg ladder in Fig.~\ref{fig:unitcell}(a) is decomposed, for $K=0$, in a diamond chain with renormalized hopping constants ($t_i^\prime=\sqrt{2}t_i$) and two decoupled sites.
Vertical dashed lines labeled 1 and 2 indicate the two inversion axes for this choice of unit cell.
(b) Energy spectrum of the diamond chain in (a) as a function of the relative momentum $k$ for $(t_1^\prime,t_2^\prime)\to \sqrt{2}t_1(1,0.3)$.
There are two degenerate flat bands around $E/t_1=0$.
Pairs of values $\alpha(\beta)$, with $\alpha,\beta=0,\pi$, indicate the Zak's phase of the corresponding band calculated from (\ref{eq:generalmodzak}) and using inversion axis 1 and 2 in (a), respectively, \textit{after} lifting the degeneracy of the dispersive bands at $k=\pi$ through the introduction of perturbative potentials at ``$+_1$'' and ``$+_2$'' sites (to preserve inversion symmetry).
The sum of the Zak's phases of the degenerate flat bands is zero independently of the inversion axis considered.
} 
\label{fig:unitcellk0}
\end{figure}

Returning to the case of spinless fermions with NN interactions, 
when $K=0$ the hopping constants of the two-leg ladder in Fig.~\ref{fig:unitcell}(a) all have the same sign.
If we perform the same basis rotation as before, given by (\ref{eq:basisrot1}) and (\ref{eq:basisrot2}), the bulk unit cell of our system becomes a diamond chain plus two decoupled sites, as depicted in Fig.~\ref{fig:unitcellk0}(a).
The two inversion axes of the diamond chain (about the plus sites) are not at the center of the unit cell, as was the case with the $t_1t_1t_2t_2$ model for $K=\pi$.
This model of the diamond chain is peculiar, since it is topologically non-trivial for both dimerizations and has topological edge states regardless of the unit cell considered (for an open chain with an \textit{integer} number of unit cells).
Looking at the unit cell of the diamond chain in Fig.~\ref{fig:unitcellk0}(a), one sees that a reversal in the dimerization is equivalent to a vertical flip ($\pi$-rotation) of the unit cell, which does not affect the topological structure of the model since all possible edge configurations for the case of OBC are equivalent under a dimerization reversal.
Additionally, this equivalence between dimerizations further translates into a topological equivalence between all four possible choices of unit cell.

The energy spectrum of the periodic diamond chain is shown in Fig.~\ref{fig:unitcellk0}(b).
The Zak's phase calculated according to (\ref{eq:generalmodzak}) is indicated for each dispersive band and for each inversion axis.
As explained above, the effect of changing from one inversion axis to the other, in the calculation of the Zak's phase, is equivalent to a dimerization reversal for a fixed inversion axis.
Therefore one sees that, regardless of the dimerization considered, an open diamond chain with an integer number of unit cells is always topologically non-trivial for an in-gap Fermi level, as
the total Zak's phase of the two lowest energy bands is $\pi$ and the flat bands have the same Zak's phase \footnote{Note that there is a basis rotation that could be performed on the diamond chain through which it would be further mapped into an SSH chain with dangling sites and renormalized hopping constants, with a similar shape to that of Fig.~1(e) in Ref.~[\onlinecite{Xiao2017}]. However, the problem loses its inversion symmetry in this modified SSH chain due to the dangling sites, and therefore our analyses, which relies on the existence of inversion symmetry in order to have $\pi$-quantized values for the Zak's phases, would not be possible}.
In this diamond chain, there is one topological state in each gap localized at the edges that end with a single site [either a $+_1$ or a $+_2$ site in Fig.~\ref{fig:unitcellk0}(a)].
For any $N$ in our original problem, the mapped diamond chain ends with two sites at both edges (either C and D or G and H sites) with on-site potential $V$ [as can be checked by mapping the $N=8$ case of Fig.~\ref{fig:2dmapping}(c) into the diamond chain], that is, the number of unit cells is fractional.
Using the reasoning followed before for $K=\pi$, the effect of a strong $V$ is to isolate the edge sites, making the chain effectively end with single sites at both inner edges, which is the condition for the presence of topological edge states.
This way, doubly degenerate topological states appear with energies inside each of the gaps in Fig.~\ref{fig:unitcellk0}(b) (see the topological in-gap energy levels for $K=0$ in Fig.~\ref{fig:energydispersion}).

The absence of topological bound states for $K=0$, when one drops the NN interaction term and considers instead Hubbard interactions, follows from the same arguments exposed in Section~\ref{sec:hubbard} for $K=\pi$ both for the opposite spins and bosonic cases, that is, when $U\to\infty$ the terminations of the longer inner diamond chain do not support edge states and, when $U=0$ (the non-interacting bosonic case), the renormalization of the edge hopping constants is responsible for driving the energy of the edge states into the bulk energy bands.

\section{Conclusions}
\label{sec:conclusions}

In this paper we have studied two-particle states in periodic Su-Schrieffer-Heeger chains with nearest-neighbor and Hubbard interactions.
This problem of two particles in a one-dimensional system was mapped into a problem of a single particle in a two dimensional lattice.
In the limit of strong nearest-neighbor interactions, for the case of two spinless fermions, the energy spectrum as a function of the center-of-mass momentum was found to have, apart from the expected energy bands of doublon states, two extra in-gap energy bands of localized states, absent when the nearest-neighbor interaction is either suppressed or substituted with a Hubbard interaction for the cases of opposite spins or identical bosons. The states of these extra bands were identified as two-body topological states that exhibit localized behavior at the edges of the internal coordinate, namely the distance between the two particles, while retaining an extended behavior over the external spatial coordinate of the periodic chain. 
Since this internal coordinate is specific to many-body systems, these topological bound states are truly many-body, in the sense that they have no single-particle counterpart.

Optical lattices appear as natural candidates for the observation of the topological bound states described here. 
Upon loading two ultracold bosonic atoms into an optical lattice with tunable hopping constants $t_i$ and repulsive on-site interaction $U$, the experimental creation and manipulation of bound states in this Bose-Hubbard model has been realized \cite{Wrinkler2006,Preiss2015,Mukherjee2016}.
To detect these topological bound states, the introduction of a nearest-neighbor interaction term $V$ in an optical lattice loaded with either bosonic or fermionic atoms is required \cite{Vargas2016}, which has proven to be experimentally more challenging than the introduction of the $U$ term.
Additionally, one has to be in the $V\gg t_1,t_2$ limit for any of the two-particle systems considered in this paper (spinless fermions, opposite spins and identical bosons, with arbitrary values for $U$ in the two latter cases).
The implementation of the extended Hubbard model on an optical lattice, with both on-site and off-site (in particular nearest-neighbor) interactions, has been achieved recently \cite{Baier2016}, which strongly suggests the possibility of a near future detection of the topological bound states studied in this paper.

Our work can be extended to $n$-particle states, with $n>2$, living in $d$ dimensions. Using the techniques detailed here, this problem can be mapped into a problem of a single-particle living in an $nd$ dimensional lattice. The effect of interactions on this
mapping, while being qualitatively the same (that is, they still map into local potentials), becomes more complex, with the appearance of multiple potential walls at sites belonging
to different $l$-dimensional hypersurfaces, with $1\leq l\leq nd-1$. As the dimensionality of the problem is increased, several energy subspaces are expected to be present in the large interaction limit,
and intrinsically many-body topological states of the kind shown here may be available in some of these subspaces.

\section*{Acknowledgments}\label{sec:acknowledments}

This work is funded by FEDER funds through the COMPETE 2020 Programme and National Funds throught FCT - Portuguese Foundation for Science and Technology under the project UID/CTM/50025/2013.
AMM acknowledges the financial support from the FCT through the grant SFRH/PD/BD/108663/2015 and from the Portuguese Institute for Nanostructures, Nanomodelling and Nanofabrication (i3N) through the grant BI/UI96/6376/2018.
RGD thanks the support by the Beijing CSRC.

\bibliography{topotwoparticles}

\begin{thebibliography}{70}%
\makeatletter
\providecommand \@ifxundefined [1]{%
 \@ifx{#1\undefined}
}%
\providecommand \@ifnum [1]{%
 \ifnum #1\expandafter \@firstoftwo
 \else \expandafter \@secondoftwo
 \fi
}%
\providecommand \@ifx [1]{%
 \ifx #1\expandafter \@firstoftwo
 \else \expandafter \@secondoftwo
 \fi
}%
\providecommand \natexlab [1]{#1}%
\providecommand \enquote  [1]{``#1''}%
\providecommand \bibnamefont  [1]{#1}%
\providecommand \bibfnamefont [1]{#1}%
\providecommand \citenamefont [1]{#1}%
\providecommand \href@noop [0]{\@secondoftwo}%
\providecommand \href [0]{\begingroup \@sanitize@url \@href}%
\providecommand \@href[1]{\@@startlink{#1}\@@href}%
\providecommand \@@href[1]{\endgroup#1\@@endlink}%
\providecommand \@sanitize@url [0]{\catcode `\\12\catcode `\$12\catcode
  `\&12\catcode `\#12\catcode `\^12\catcode `\_12\catcode `\%12\relax}%
\providecommand \@@startlink[1]{}%
\providecommand \@@endlink[0]{}%
\providecommand \url  [0]{\begingroup\@sanitize@url \@url }%
\providecommand \@url [1]{\endgroup\@href {#1}{\urlprefix }}%
\providecommand \urlprefix  [0]{URL }%
\providecommand \Eprint [0]{\href }%
\providecommand \doibase [0]{http://dx.doi.org/}%
\providecommand \selectlanguage [0]{\@gobble}%
\providecommand \bibinfo  [0]{\@secondoftwo}%
\providecommand \bibfield  [0]{\@secondoftwo}%
\providecommand \translation [1]{[#1]}%
\providecommand \BibitemOpen [0]{}%
\providecommand \bibitemStop [0]{}%
\providecommand \bibitemNoStop [0]{.\EOS\space}%
\providecommand \EOS [0]{\spacefactor3000\relax}%
\providecommand \BibitemShut  [1]{\csname bibitem#1\endcsname}%
\let\auto@bib@innerbib\@empty
\bibitem [{\citenamefont {Niu}\ and\ \citenamefont {Thouless}(1984)}]{Niu1984}%
  \BibitemOpen
  \bibfield  {author} {\bibinfo {author} {\bibfnamefont {Q.}~\bibnamefont
  {Niu}}\ and\ \bibinfo {author} {\bibfnamefont {D.~J.}\ \bibnamefont
  {Thouless}},\ }\href {http://stacks.iop.org/0305-4470/17/i=12/a=016}
  {\bibfield  {journal} {\bibinfo  {journal} {J. Phys. A: Math. Gen.}\ }\textbf
  {\bibinfo {volume} {17}},\ \bibinfo {pages} {2453} (\bibinfo {year}
  {1984})}\BibitemShut {NoStop}%
\bibitem [{\citenamefont {Guo}\ and\ \citenamefont {Shen}(2011)}]{Guo2011}%
  \BibitemOpen
  \bibfield  {author} {\bibinfo {author} {\bibfnamefont {H.}~\bibnamefont
  {Guo}}\ and\ \bibinfo {author} {\bibfnamefont {S.-Q.}\ \bibnamefont {Shen}},\
  }\href {\doibase 10.1103/PhysRevB.84.195107} {\bibfield  {journal} {\bibinfo
  {journal} {Phys. Rev. B}\ }\textbf {\bibinfo {volume} {84}},\ \bibinfo
  {pages} {195107} (\bibinfo {year} {2011})}\BibitemShut {NoStop}%
\bibitem [{\citenamefont {Grusdt}\ \emph {et~al.}(2013)\citenamefont {Grusdt},
  \citenamefont {H\"oning},\ and\ \citenamefont {Fleischhauer}}]{Grusdt2013}%
  \BibitemOpen
  \bibfield  {author} {\bibinfo {author} {\bibfnamefont {F.}~\bibnamefont
  {Grusdt}}, \bibinfo {author} {\bibfnamefont {M.}~\bibnamefont {H\"oning}}, \
  and\ \bibinfo {author} {\bibfnamefont {M.}~\bibnamefont {Fleischhauer}},\
  }\href {\doibase 10.1103/PhysRevLett.110.260405} {\bibfield  {journal}
  {\bibinfo  {journal} {Phys. Rev. Lett.}\ }\textbf {\bibinfo {volume} {110}},\
  \bibinfo {pages} {260405} (\bibinfo {year} {2013})}\BibitemShut {NoStop}%
\bibitem [{\citenamefont {Li}\ and\ \citenamefont {Chen}(2015)}]{Li2015}%
  \BibitemOpen
  \bibfield  {author} {\bibinfo {author} {\bibfnamefont {L.}~\bibnamefont
  {Li}}\ and\ \bibinfo {author} {\bibfnamefont {S.}~\bibnamefont {Chen}},\
  }\href {http://stacks.iop.org/0295-5075/109/i=4/a=40006} {\bibfield
  {journal} {\bibinfo  {journal} {EPL}\ }\textbf {\bibinfo {volume} {109}},\
  \bibinfo {pages} {40006} (\bibinfo {year} {2015})}\BibitemShut {NoStop}%
\bibitem [{\citenamefont {Vanhala}\ \emph {et~al.}(2016)\citenamefont
  {Vanhala}, \citenamefont {Siro}, \citenamefont {Liang}, \citenamefont
  {Troyer}, \citenamefont {Harju},\ and\ \citenamefont
  {T\"orm\"a}}]{Vanhala2016}%
  \BibitemOpen
  \bibfield  {author} {\bibinfo {author} {\bibfnamefont {T.~I.}\ \bibnamefont
  {Vanhala}}, \bibinfo {author} {\bibfnamefont {T.}~\bibnamefont {Siro}},
  \bibinfo {author} {\bibfnamefont {L.}~\bibnamefont {Liang}}, \bibinfo
  {author} {\bibfnamefont {M.}~\bibnamefont {Troyer}}, \bibinfo {author}
  {\bibfnamefont {A.}~\bibnamefont {Harju}}, \ and\ \bibinfo {author}
  {\bibfnamefont {P.}~\bibnamefont {T\"orm\"a}},\ }\href {\doibase
  10.1103/PhysRevLett.116.225305} {\bibfield  {journal} {\bibinfo  {journal}
  {Phys. Rev. Lett.}\ }\textbf {\bibinfo {volume} {116}},\ \bibinfo {pages}
  {225305} (\bibinfo {year} {2016})}\BibitemShut {NoStop}%
\bibitem [{\citenamefont {Kumar}\ \emph {et~al.}(2016)\citenamefont {Kumar},
  \citenamefont {Mertz},\ and\ \citenamefont {Hofstetter}}]{Kumar2016}%
  \BibitemOpen
  \bibfield  {author} {\bibinfo {author} {\bibfnamefont {P.}~\bibnamefont
  {Kumar}}, \bibinfo {author} {\bibfnamefont {T.}~\bibnamefont {Mertz}}, \ and\
  \bibinfo {author} {\bibfnamefont {W.}~\bibnamefont {Hofstetter}},\ }\href
  {\doibase 10.1103/PhysRevB.94.115161} {\bibfield  {journal} {\bibinfo
  {journal} {Phys. Rev. B}\ }\textbf {\bibinfo {volume} {94}},\ \bibinfo
  {pages} {115161} (\bibinfo {year} {2016})}\BibitemShut {NoStop}%
\bibitem [{\citenamefont {Guo}\ \emph {et~al.}(2012)\citenamefont {Guo},
  \citenamefont {Shen},\ and\ \citenamefont {Feng}}]{Guo2012}%
  \BibitemOpen
  \bibfield  {author} {\bibinfo {author} {\bibfnamefont {H.}~\bibnamefont
  {Guo}}, \bibinfo {author} {\bibfnamefont {S.-Q.}\ \bibnamefont {Shen}}, \
  and\ \bibinfo {author} {\bibfnamefont {S.}~\bibnamefont {Feng}},\ }\href
  {\doibase 10.1103/PhysRevB.86.085124} {\bibfield  {journal} {\bibinfo
  {journal} {Phys. Rev. B}\ }\textbf {\bibinfo {volume} {86}},\ \bibinfo
  {pages} {085124} (\bibinfo {year} {2012})}\BibitemShut {NoStop}%
\bibitem [{\citenamefont {Budich}\ and\ \citenamefont
  {Ardonne}(2013)}]{Budich2013}%
  \BibitemOpen
  \bibfield  {author} {\bibinfo {author} {\bibfnamefont {J.~C.}\ \bibnamefont
  {Budich}}\ and\ \bibinfo {author} {\bibfnamefont {E.}~\bibnamefont
  {Ardonne}},\ }\href {\doibase 10.1103/PhysRevB.88.035139} {\bibfield
  {journal} {\bibinfo  {journal} {Phys. Rev. B}\ }\textbf {\bibinfo {volume}
  {88}},\ \bibinfo {pages} {035139} (\bibinfo {year} {2013})}\BibitemShut
  {NoStop}%
\bibitem [{\citenamefont {Kivelson}\ and\ \citenamefont
  {Heim}(1982)}]{Kivelson1982}%
  \BibitemOpen
  \bibfield  {author} {\bibinfo {author} {\bibfnamefont {S.}~\bibnamefont
  {Kivelson}}\ and\ \bibinfo {author} {\bibfnamefont {D.~E.}\ \bibnamefont
  {Heim}},\ }\href {\doibase 10.1103/PhysRevB.26.4278} {\bibfield  {journal}
  {\bibinfo  {journal} {Phys. Rev. B}\ }\textbf {\bibinfo {volume} {26}},\
  \bibinfo {pages} {4278} (\bibinfo {year} {1982})}\BibitemShut {NoStop}%
\bibitem [{\citenamefont {Liu}\ \emph {et~al.}(2013)\citenamefont {Liu},
  \citenamefont {Liu},\ and\ \citenamefont {Cheng}}]{Liu2013}%
  \BibitemOpen
  \bibfield  {author} {\bibinfo {author} {\bibfnamefont {X.-J.}\ \bibnamefont
  {Liu}}, \bibinfo {author} {\bibfnamefont {Z.-X.}\ \bibnamefont {Liu}}, \ and\
  \bibinfo {author} {\bibfnamefont {M.}~\bibnamefont {Cheng}},\ }\href
  {\doibase 10.1103/PhysRevLett.110.076401} {\bibfield  {journal} {\bibinfo
  {journal} {Phys. Rev. Lett.}\ }\textbf {\bibinfo {volume} {110}},\ \bibinfo
  {pages} {076401} (\bibinfo {year} {2013})}\BibitemShut {NoStop}%
\bibitem [{\citenamefont {Weber}\ \emph {et~al.}(2015)\citenamefont {Weber},
  \citenamefont {Assaad},\ and\ \citenamefont {Hohenadler}}]{Weber2015}%
  \BibitemOpen
  \bibfield  {author} {\bibinfo {author} {\bibfnamefont {M.}~\bibnamefont
  {Weber}}, \bibinfo {author} {\bibfnamefont {F.~F.}\ \bibnamefont {Assaad}}, \
  and\ \bibinfo {author} {\bibfnamefont {M.}~\bibnamefont {Hohenadler}},\
  }\href {\doibase 10.1103/PhysRevB.91.245147} {\bibfield  {journal} {\bibinfo
  {journal} {Phys. Rev. B}\ }\textbf {\bibinfo {volume} {91}},\ \bibinfo
  {pages} {245147} (\bibinfo {year} {2015})}\BibitemShut {NoStop}%
\bibitem [{\citenamefont {Zhang}\ \emph {et~al.}(2013)\citenamefont {Zhang},
  \citenamefont {Braak},\ and\ \citenamefont {Kollar}}]{Zhang2013}%
  \BibitemOpen
  \bibfield  {author} {\bibinfo {author} {\bibfnamefont {J.~M.}\ \bibnamefont
  {Zhang}}, \bibinfo {author} {\bibfnamefont {D.}~\bibnamefont {Braak}}, \ and\
  \bibinfo {author} {\bibfnamefont {M.}~\bibnamefont {Kollar}},\ }\href
  {\doibase 10.1103/PhysRevA.87.023613} {\bibfield  {journal} {\bibinfo
  {journal} {Phys. Rev. A}\ }\textbf {\bibinfo {volume} {87}},\ \bibinfo
  {pages} {023613} (\bibinfo {year} {2013})}\BibitemShut {NoStop}%
\bibitem [{\citenamefont {Bello}\ \emph {et~al.}(2016)\citenamefont {Bello},
  \citenamefont {Creffield},\ and\ \citenamefont {Platero}}]{Bello2016}%
  \BibitemOpen
  \bibfield  {author} {\bibinfo {author} {\bibfnamefont {M.}~\bibnamefont
  {Bello}}, \bibinfo {author} {\bibfnamefont {C.~E.}\ \bibnamefont
  {Creffield}}, \ and\ \bibinfo {author} {\bibfnamefont {G.}~\bibnamefont
  {Platero}},\ }\href {http://dx.doi.org/10.1038/srep22562} {\bibfield
  {journal} {\bibinfo  {journal} {Sci. Rep.}\ }\textbf {\bibinfo {volume}
  {6}},\ \bibinfo {pages} {22562} (\bibinfo {year} {2016})}\BibitemShut
  {NoStop}%
\bibitem [{\citenamefont {Di~Liberto}\ \emph {et~al.}(2016)\citenamefont
  {Di~Liberto}, \citenamefont {Recati}, \citenamefont {Carusotto},\ and\
  \citenamefont {Menotti}}]{Liberto2016}%
  \BibitemOpen
  \bibfield  {author} {\bibinfo {author} {\bibfnamefont {M.}~\bibnamefont
  {Di~Liberto}}, \bibinfo {author} {\bibfnamefont {A.}~\bibnamefont {Recati}},
  \bibinfo {author} {\bibfnamefont {I.}~\bibnamefont {Carusotto}}, \ and\
  \bibinfo {author} {\bibfnamefont {C.}~\bibnamefont {Menotti}},\ }\href
  {\doibase 10.1103/PhysRevA.94.062704} {\bibfield  {journal} {\bibinfo
  {journal} {Phys. Rev. A}\ }\textbf {\bibinfo {volume} {94}},\ \bibinfo
  {pages} {062704} (\bibinfo {year} {2016})}\BibitemShut {NoStop}%
\bibitem [{\citenamefont {Gorlach}\ and\ \citenamefont
  {Poddubny}(2017{\natexlab{a}})}]{Gorlach2017b}%
  \BibitemOpen
  \bibfield  {author} {\bibinfo {author} {\bibfnamefont {M.~A.}\ \bibnamefont
  {Gorlach}}\ and\ \bibinfo {author} {\bibfnamefont {A.~N.}\ \bibnamefont
  {Poddubny}},\ }\href {\doibase 10.1103/PhysRevA.95.053866} {\bibfield
  {journal} {\bibinfo  {journal} {Phys. Rev. A}\ }\textbf {\bibinfo {volume}
  {95}},\ \bibinfo {pages} {053866} (\bibinfo {year}
  {2017}{\natexlab{a}})}\BibitemShut {NoStop}%
\bibitem [{\citenamefont {Scott}\ \emph {et~al.}(1994)\citenamefont {Scott},
  \citenamefont {Eilbeck},\ and\ \citenamefont {Gilh{\o}j}}]{Scott1994}%
  \BibitemOpen
  \bibfield  {author} {\bibinfo {author} {\bibfnamefont {A.}~\bibnamefont
  {Scott}}, \bibinfo {author} {\bibfnamefont {J.}~\bibnamefont {Eilbeck}}, \
  and\ \bibinfo {author} {\bibfnamefont {H.}~\bibnamefont {Gilh{\o}j}},\ }\href
  {\doibase https://doi.org/10.1016/0167-2789(94)90115-5} {\bibfield  {journal}
  {\bibinfo  {journal} {Physica D}\ }\textbf {\bibinfo {volume} {78}},\
  \bibinfo {pages} {194 } (\bibinfo {year} {1994})}\BibitemShut {NoStop}%
\bibitem [{\citenamefont {Creffield}\ and\ \citenamefont
  {Platero}(2004)}]{Creffield2004}%
  \BibitemOpen
  \bibfield  {author} {\bibinfo {author} {\bibfnamefont {C.~E.}\ \bibnamefont
  {Creffield}}\ and\ \bibinfo {author} {\bibfnamefont {G.}~\bibnamefont
  {Platero}},\ }\href {\doibase 10.1103/PhysRevB.69.165312} {\bibfield
  {journal} {\bibinfo  {journal} {Phys. Rev. B}\ }\textbf {\bibinfo {volume}
  {69}},\ \bibinfo {pages} {165312} (\bibinfo {year} {2004})}\BibitemShut
  {NoStop}%
\bibitem [{\citenamefont {Valiente}\ and\ \citenamefont
  {Petrosyan}(2008)}]{Valiente2008}%
  \BibitemOpen
  \bibfield  {author} {\bibinfo {author} {\bibfnamefont {M.}~\bibnamefont
  {Valiente}}\ and\ \bibinfo {author} {\bibfnamefont {D.}~\bibnamefont
  {Petrosyan}},\ }\href {http://stacks.iop.org/0953-4075/41/i=16/a=161002}
  {\bibfield  {journal} {\bibinfo  {journal} {J. Phys. B}\ }\textbf {\bibinfo
  {volume} {41}},\ \bibinfo {pages} {161002} (\bibinfo {year}
  {2008})}\BibitemShut {NoStop}%
\bibitem [{\citenamefont {Piil}\ \emph {et~al.}(2008)\citenamefont {Piil},
  \citenamefont {Nygaard},\ and\ \citenamefont {M\o{}lmer}}]{Piil2008}%
  \BibitemOpen
  \bibfield  {author} {\bibinfo {author} {\bibfnamefont {R.~T.}\ \bibnamefont
  {Piil}}, \bibinfo {author} {\bibfnamefont {N.}~\bibnamefont {Nygaard}}, \
  and\ \bibinfo {author} {\bibfnamefont {K.}~\bibnamefont {M\o{}lmer}},\ }\href
  {\doibase 10.1103/PhysRevA.78.033611} {\bibfield  {journal} {\bibinfo
  {journal} {Phys. Rev. A}\ }\textbf {\bibinfo {volume} {78}},\ \bibinfo
  {pages} {033611} (\bibinfo {year} {2008})}\BibitemShut {NoStop}%
\bibitem [{\citenamefont {Valiente}\ and\ \citenamefont
  {Petrosyan}(2009)}]{Valiente2009}%
  \BibitemOpen
  \bibfield  {author} {\bibinfo {author} {\bibfnamefont {M.}~\bibnamefont
  {Valiente}}\ and\ \bibinfo {author} {\bibfnamefont {D.}~\bibnamefont
  {Petrosyan}},\ }\href {http://stacks.iop.org/0953-4075/42/i=12/a=121001}
  {\bibfield  {journal} {\bibinfo  {journal} {J. Phys. B}\ }\textbf {\bibinfo
  {volume} {42}},\ \bibinfo {pages} {121001} (\bibinfo {year}
  {2009})}\BibitemShut {NoStop}%
\bibitem [{\citenamefont {Javanainen}\ \emph {et~al.}(2010)\citenamefont
  {Javanainen}, \citenamefont {Odong},\ and\ \citenamefont
  {Sanders}}]{Javanainen2010}%
  \BibitemOpen
  \bibfield  {author} {\bibinfo {author} {\bibfnamefont {J.}~\bibnamefont
  {Javanainen}}, \bibinfo {author} {\bibfnamefont {O.}~\bibnamefont {Odong}}, \
  and\ \bibinfo {author} {\bibfnamefont {J.~C.}\ \bibnamefont {Sanders}},\
  }\href {\doibase 10.1103/PhysRevA.81.043609} {\bibfield  {journal} {\bibinfo
  {journal} {Phys. Rev. A}\ }\textbf {\bibinfo {volume} {81}},\ \bibinfo
  {pages} {043609} (\bibinfo {year} {2010})}\BibitemShut {NoStop}%
\bibitem [{\citenamefont {Wang}\ and\ \citenamefont {Liang}(2010)}]{Wang2010}%
  \BibitemOpen
  \bibfield  {author} {\bibinfo {author} {\bibfnamefont {Y.-M.}\ \bibnamefont
  {Wang}}\ and\ \bibinfo {author} {\bibfnamefont {J.-Q.}\ \bibnamefont
  {Liang}},\ }\href {\doibase 10.1103/PhysRevA.81.045601} {\bibfield  {journal}
  {\bibinfo  {journal} {Phys. Rev. A}\ }\textbf {\bibinfo {volume} {81}},\
  \bibinfo {pages} {045601} (\bibinfo {year} {2010})}\BibitemShut {NoStop}%
\bibitem [{\citenamefont {Creffield}\ and\ \citenamefont
  {Platero}(2010)}]{Creffield2010}%
  \BibitemOpen
  \bibfield  {author} {\bibinfo {author} {\bibfnamefont {C.~E.}\ \bibnamefont
  {Creffield}}\ and\ \bibinfo {author} {\bibfnamefont {G.}~\bibnamefont
  {Platero}},\ }\href {\doibase 10.1103/PhysRevLett.105.086804} {\bibfield
  {journal} {\bibinfo  {journal} {Phys. Rev. Lett.}\ }\textbf {\bibinfo
  {volume} {105}},\ \bibinfo {pages} {086804} (\bibinfo {year}
  {2010})}\BibitemShut {NoStop}%
\bibitem [{\citenamefont {Valiente}\ \emph {et~al.}(2010)\citenamefont
  {Valiente}, \citenamefont {Küster},\ and\ \citenamefont
  {Saenz}}]{Valiente2010}%
  \BibitemOpen
  \bibfield  {author} {\bibinfo {author} {\bibfnamefont {M.}~\bibnamefont
  {Valiente}}, \bibinfo {author} {\bibfnamefont {M.}~\bibnamefont {Küster}}, \
  and\ \bibinfo {author} {\bibfnamefont {A.}~\bibnamefont {Saenz}},\ }\href
  {http://stacks.iop.org/0295-5075/92/i=1/a=10001} {\bibfield  {journal}
  {\bibinfo  {journal} {EPL}\ }\textbf {\bibinfo {volume} {92}},\ \bibinfo
  {pages} {10001} (\bibinfo {year} {2010})}\BibitemShut {NoStop}%
\bibitem [{\citenamefont {Nguenang}\ \emph {et~al.}(2012)\citenamefont
  {Nguenang}, \citenamefont {Flach},\ and\ \citenamefont
  {Khomeriki}}]{Nguenang2012}%
  \BibitemOpen
  \bibfield  {author} {\bibinfo {author} {\bibfnamefont {J.-P.}\ \bibnamefont
  {Nguenang}}, \bibinfo {author} {\bibfnamefont {S.}~\bibnamefont {Flach}}, \
  and\ \bibinfo {author} {\bibfnamefont {R.}~\bibnamefont {Khomeriki}},\ }\href
  {\doibase http://dx.doi.org/10.1016/j.physleta.2011.11.048} {\bibfield
  {journal} {\bibinfo  {journal} {Phys. Lett. A}\ }\textbf {\bibinfo {volume}
  {376}},\ \bibinfo {pages} {472 } (\bibinfo {year} {2012})}\BibitemShut
  {NoStop}%
\bibitem [{\citenamefont {Longhi}\ and\ \citenamefont
  {Della~Valle}(2013)}]{Longhi2013}%
  \BibitemOpen
  \bibfield  {author} {\bibinfo {author} {\bibfnamefont {S.}~\bibnamefont
  {Longhi}}\ and\ \bibinfo {author} {\bibfnamefont {G.}~\bibnamefont
  {Della~Valle}},\ }\href {\doibase 10.1103/PhysRevA.87.013634} {\bibfield
  {journal} {\bibinfo  {journal} {Phys. Rev. A}\ }\textbf {\bibinfo {volume}
  {87}},\ \bibinfo {pages} {013634} (\bibinfo {year} {2013})}\BibitemShut
  {NoStop}%
\bibitem [{\citenamefont {Qin}\ \emph {et~al.}(2014)\citenamefont {Qin},
  \citenamefont {Ke}, \citenamefont {Guan}, \citenamefont {Li}, \citenamefont
  {Andrei},\ and\ \citenamefont {Lee}}]{Qin2014}%
  \BibitemOpen
  \bibfield  {author} {\bibinfo {author} {\bibfnamefont {X.}~\bibnamefont
  {Qin}}, \bibinfo {author} {\bibfnamefont {Y.}~\bibnamefont {Ke}}, \bibinfo
  {author} {\bibfnamefont {X.}~\bibnamefont {Guan}}, \bibinfo {author}
  {\bibfnamefont {Z.}~\bibnamefont {Li}}, \bibinfo {author} {\bibfnamefont
  {N.}~\bibnamefont {Andrei}}, \ and\ \bibinfo {author} {\bibfnamefont
  {C.}~\bibnamefont {Lee}},\ }\href {\doibase 10.1103/PhysRevA.90.062301}
  {\bibfield  {journal} {\bibinfo  {journal} {Phys. Rev. A}\ }\textbf {\bibinfo
  {volume} {90}},\ \bibinfo {pages} {062301} (\bibinfo {year}
  {2014})}\BibitemShut {NoStop}%
\bibitem [{\citenamefont {Barbiero}\ \emph {et~al.}(2015)\citenamefont
  {Barbiero}, \citenamefont {Menotti}, \citenamefont {Recati},\ and\
  \citenamefont {Santos}}]{Barbiero2015}%
  \BibitemOpen
  \bibfield  {author} {\bibinfo {author} {\bibfnamefont {L.}~\bibnamefont
  {Barbiero}}, \bibinfo {author} {\bibfnamefont {C.}~\bibnamefont {Menotti}},
  \bibinfo {author} {\bibfnamefont {A.}~\bibnamefont {Recati}}, \ and\ \bibinfo
  {author} {\bibfnamefont {L.}~\bibnamefont {Santos}},\ }\href {\doibase
  10.1103/PhysRevB.92.180406} {\bibfield  {journal} {\bibinfo  {journal} {Phys.
  Rev. B}\ }\textbf {\bibinfo {volume} {92}},\ \bibinfo {pages} {180406}
  (\bibinfo {year} {2015})}\BibitemShut {NoStop}%
\bibitem [{\citenamefont {Winkler}\ \emph {et~al.}(2006)\citenamefont
  {Winkler}, \citenamefont {Thalhammer}, \citenamefont {Lang}, \citenamefont
  {Grimm}, \citenamefont {Hecker~Denschlag}, \citenamefont {Daley},
  \citenamefont {Kantian}, \citenamefont {Büchler},\ and\ \citenamefont
  {Zoller}}]{Wrinkler2006}%
  \BibitemOpen
  \bibfield  {author} {\bibinfo {author} {\bibfnamefont {K.}~\bibnamefont
  {Winkler}}, \bibinfo {author} {\bibfnamefont {G.}~\bibnamefont {Thalhammer}},
  \bibinfo {author} {\bibfnamefont {F.}~\bibnamefont {Lang}}, \bibinfo {author}
  {\bibfnamefont {R.}~\bibnamefont {Grimm}}, \bibinfo {author} {\bibfnamefont
  {J.}~\bibnamefont {Hecker~Denschlag}}, \bibinfo {author} {\bibfnamefont
  {A.~J.}\ \bibnamefont {Daley}}, \bibinfo {author} {\bibfnamefont
  {A.}~\bibnamefont {Kantian}}, \bibinfo {author} {\bibfnamefont {H.~P.}\
  \bibnamefont {Büchler}}, \ and\ \bibinfo {author} {\bibfnamefont
  {P.}~\bibnamefont {Zoller}},\ }\href {http://dx.doi.org/10.1038/nature04918}
  {\bibfield  {journal} {\bibinfo  {journal} {Nature}\ }\textbf {\bibinfo
  {volume} {441}},\ \bibinfo {pages} {853} (\bibinfo {year}
  {2006})}\BibitemShut {NoStop}%
\bibitem [{\citenamefont {Folling}\ \emph {et~al.}(2007)\citenamefont
  {Folling}, \citenamefont {Trotzky}, \citenamefont {Cheinet}, \citenamefont
  {Feld}, \citenamefont {Saers}, \citenamefont {Widera}, \citenamefont
  {Muller},\ and\ \citenamefont {Bloch}}]{Folling2007}%
  \BibitemOpen
  \bibfield  {author} {\bibinfo {author} {\bibfnamefont {S.}~\bibnamefont
  {Folling}}, \bibinfo {author} {\bibfnamefont {S.}~\bibnamefont {Trotzky}},
  \bibinfo {author} {\bibfnamefont {P.}~\bibnamefont {Cheinet}}, \bibinfo
  {author} {\bibfnamefont {M.}~\bibnamefont {Feld}}, \bibinfo {author}
  {\bibfnamefont {R.}~\bibnamefont {Saers}}, \bibinfo {author} {\bibfnamefont
  {A.}~\bibnamefont {Widera}}, \bibinfo {author} {\bibfnamefont
  {T.}~\bibnamefont {Muller}}, \ and\ \bibinfo {author} {\bibfnamefont
  {I.}~\bibnamefont {Bloch}},\ }\href {http://dx.doi.org/10.1038/nature06112}
  {\bibfield  {journal} {\bibinfo  {journal} {Nature}\ }\textbf {\bibinfo
  {volume} {448}},\ \bibinfo {pages} {1029} (\bibinfo {year}
  {2007})}\BibitemShut {NoStop}%
\bibitem [{\citenamefont {Longhi}(2011)}]{Longhi2011}%
  \BibitemOpen
  \bibfield  {author} {\bibinfo {author} {\bibfnamefont {S.}~\bibnamefont
  {Longhi}},\ }\href {\doibase 10.1364/OL.36.003248} {\bibfield  {journal}
  {\bibinfo  {journal} {Opt. Lett.}\ }\textbf {\bibinfo {volume} {36}},\
  \bibinfo {pages} {3248} (\bibinfo {year} {2011})}\BibitemShut {NoStop}%
\bibitem [{\citenamefont {Krimer}\ and\ \citenamefont
  {Khomeriki}(2011)}]{Krimer2011}%
  \BibitemOpen
  \bibfield  {author} {\bibinfo {author} {\bibfnamefont {D.~O.}\ \bibnamefont
  {Krimer}}\ and\ \bibinfo {author} {\bibfnamefont {R.}~\bibnamefont
  {Khomeriki}},\ }\href {\doibase 10.1103/PhysRevA.84.041807} {\bibfield
  {journal} {\bibinfo  {journal} {Phys. Rev. A}\ }\textbf {\bibinfo {volume}
  {84}},\ \bibinfo {pages} {041807} (\bibinfo {year} {2011})}\BibitemShut
  {NoStop}%
\bibitem [{\citenamefont {Preiss}\ \emph {et~al.}(2015)\citenamefont {Preiss},
  \citenamefont {Ma}, \citenamefont {Tai}, \citenamefont {Lukin}, \citenamefont
  {Rispoli}, \citenamefont {Zupancic}, \citenamefont {Lahini}, \citenamefont
  {Islam},\ and\ \citenamefont {Greiner}}]{Preiss2015}%
  \BibitemOpen
  \bibfield  {author} {\bibinfo {author} {\bibfnamefont {P.~M.}\ \bibnamefont
  {Preiss}}, \bibinfo {author} {\bibfnamefont {R.}~\bibnamefont {Ma}}, \bibinfo
  {author} {\bibfnamefont {M.~E.}\ \bibnamefont {Tai}}, \bibinfo {author}
  {\bibfnamefont {A.}~\bibnamefont {Lukin}}, \bibinfo {author} {\bibfnamefont
  {M.}~\bibnamefont {Rispoli}}, \bibinfo {author} {\bibfnamefont
  {P.}~\bibnamefont {Zupancic}}, \bibinfo {author} {\bibfnamefont
  {Y.}~\bibnamefont {Lahini}}, \bibinfo {author} {\bibfnamefont
  {R.}~\bibnamefont {Islam}}, \ and\ \bibinfo {author} {\bibfnamefont
  {M.}~\bibnamefont {Greiner}},\ }\href {\doibase 10.1126/science.1260364}
  {\bibfield  {journal} {\bibinfo  {journal} {Science}\ }\textbf {\bibinfo
  {volume} {347}},\ \bibinfo {pages} {1229} (\bibinfo {year}
  {2015})}\BibitemShut {NoStop}%
\bibitem [{\citenamefont {Mukherjee}\ \emph {et~al.}(2016)\citenamefont
  {Mukherjee}, \citenamefont {Valiente}, \citenamefont {Goldman}, \citenamefont
  {Spracklen}, \citenamefont {Andersson}, \citenamefont {\"Ohberg},\ and\
  \citenamefont {Thomson}}]{Mukherjee2016}%
  \BibitemOpen
  \bibfield  {author} {\bibinfo {author} {\bibfnamefont {S.}~\bibnamefont
  {Mukherjee}}, \bibinfo {author} {\bibfnamefont {M.}~\bibnamefont {Valiente}},
  \bibinfo {author} {\bibfnamefont {N.}~\bibnamefont {Goldman}}, \bibinfo
  {author} {\bibfnamefont {A.}~\bibnamefont {Spracklen}}, \bibinfo {author}
  {\bibfnamefont {E.}~\bibnamefont {Andersson}}, \bibinfo {author}
  {\bibfnamefont {P.}~\bibnamefont {\"Ohberg}}, \ and\ \bibinfo {author}
  {\bibfnamefont {R.~R.}\ \bibnamefont {Thomson}},\ }\href {\doibase
  10.1103/PhysRevA.94.053853} {\bibfield  {journal} {\bibinfo  {journal} {Phys.
  Rev. A}\ }\textbf {\bibinfo {volume} {94}},\ \bibinfo {pages} {053853}
  (\bibinfo {year} {2016})}\BibitemShut {NoStop}%
\bibitem [{\citenamefont {Pinto}\ \emph {et~al.}(2009)\citenamefont {Pinto},
  \citenamefont {Haque},\ and\ \citenamefont {Flach}}]{Pinto2009}%
  \BibitemOpen
  \bibfield  {author} {\bibinfo {author} {\bibfnamefont {R.~A.}\ \bibnamefont
  {Pinto}}, \bibinfo {author} {\bibfnamefont {M.}~\bibnamefont {Haque}}, \ and\
  \bibinfo {author} {\bibfnamefont {S.}~\bibnamefont {Flach}},\ }\href
  {\doibase 10.1103/PhysRevA.79.052118} {\bibfield  {journal} {\bibinfo
  {journal} {Phys. Rev. A}\ }\textbf {\bibinfo {volume} {79}},\ \bibinfo
  {pages} {052118} (\bibinfo {year} {2009})}\BibitemShut {NoStop}%
\bibitem [{\citenamefont {Haque}(2010)}]{Haque2010}%
  \BibitemOpen
  \bibfield  {author} {\bibinfo {author} {\bibfnamefont {M.}~\bibnamefont
  {Haque}},\ }\href {\doibase 10.1103/PhysRevA.82.012108} {\bibfield  {journal}
  {\bibinfo  {journal} {Phys. Rev. A}\ }\textbf {\bibinfo {volume} {82}},\
  \bibinfo {pages} {012108} (\bibinfo {year} {2010})}\BibitemShut {NoStop}%
\bibitem [{\citenamefont {Banchi}\ and\ \citenamefont
  {Vaia}(2013)}]{Banchi2013}%
  \BibitemOpen
  \bibfield  {author} {\bibinfo {author} {\bibfnamefont {L.}~\bibnamefont
  {Banchi}}\ and\ \bibinfo {author} {\bibfnamefont {R.}~\bibnamefont {Vaia}},\
  }\href {\doibase 10.1063/1.4797477} {\bibfield  {journal} {\bibinfo
  {journal} {J. Math. Phys.}\ }\textbf {\bibinfo {volume} {54}},\ \bibinfo
  {pages} {043501} (\bibinfo {year} {2013})}\BibitemShut {NoStop}%
\bibitem [{\citenamefont {Compagno}\ \emph {et~al.}(2017)\citenamefont
  {Compagno}, \citenamefont {Banchi}, \citenamefont {Gross},\ and\
  \citenamefont {Bose}}]{Compagno2017}%
  \BibitemOpen
  \bibfield  {author} {\bibinfo {author} {\bibfnamefont {E.}~\bibnamefont
  {Compagno}}, \bibinfo {author} {\bibfnamefont {L.}~\bibnamefont {Banchi}},
  \bibinfo {author} {\bibfnamefont {C.}~\bibnamefont {Gross}}, \ and\ \bibinfo
  {author} {\bibfnamefont {S.}~\bibnamefont {Bose}},\ }\href {\doibase
  10.1103/PhysRevA.95.012307} {\bibfield  {journal} {\bibinfo  {journal} {Phys.
  Rev. A}\ }\textbf {\bibinfo {volume} {95}},\ \bibinfo {pages} {012307}
  (\bibinfo {year} {2017})}\BibitemShut {NoStop}%
\bibitem [{\citenamefont {Bello}\ \emph {et~al.}(2017)\citenamefont {Bello},
  \citenamefont {Creffield},\ and\ \citenamefont {Platero}}]{Bello2017}%
  \BibitemOpen
  \bibfield  {author} {\bibinfo {author} {\bibfnamefont {M.}~\bibnamefont
  {Bello}}, \bibinfo {author} {\bibfnamefont {C.~E.}\ \bibnamefont
  {Creffield}}, \ and\ \bibinfo {author} {\bibfnamefont {G.}~\bibnamefont
  {Platero}},\ }\href {\doibase 10.1103/PhysRevB.95.094303} {\bibfield
  {journal} {\bibinfo  {journal} {Phys. Rev. B}\ }\textbf {\bibinfo {volume}
  {95}},\ \bibinfo {pages} {094303} (\bibinfo {year} {2017})}\BibitemShut
  {NoStop}%
\bibitem [{\citenamefont {Marques}\ and\ \citenamefont
  {Dias}(2017)}]{Marques2017}%
  \BibitemOpen
  \bibfield  {author} {\bibinfo {author} {\bibfnamefont {A.~M.}\ \bibnamefont
  {Marques}}\ and\ \bibinfo {author} {\bibfnamefont {R.~G.}\ \bibnamefont
  {Dias}},\ }\href {\doibase 10.1103/PhysRevB.95.115443} {\bibfield  {journal}
  {\bibinfo  {journal} {Phys. Rev. B}\ }\textbf {\bibinfo {volume} {95}},\
  \bibinfo {pages} {115443} (\bibinfo {year} {2017})}\BibitemShut {NoStop}%
\bibitem [{\citenamefont {Xiao}\ \emph {et~al.}(2017)\citenamefont {Xiao},
  \citenamefont {Ma}, \citenamefont {Zhang},\ and\ \citenamefont
  {Chan}}]{Xiao2017}%
  \BibitemOpen
  \bibfield  {author} {\bibinfo {author} {\bibfnamefont {Y.-X.}\ \bibnamefont
  {Xiao}}, \bibinfo {author} {\bibfnamefont {G.}~\bibnamefont {Ma}}, \bibinfo
  {author} {\bibfnamefont {Z.-Q.}\ \bibnamefont {Zhang}}, \ and\ \bibinfo
  {author} {\bibfnamefont {C.~T.}\ \bibnamefont {Chan}},\ }\href {\doibase
  10.1103/PhysRevLett.118.166803} {\bibfield  {journal} {\bibinfo  {journal}
  {Phys. Rev. Lett.}\ }\textbf {\bibinfo {volume} {118}},\ \bibinfo {pages}
  {166803} (\bibinfo {year} {2017})}\BibitemShut {NoStop}%
\bibitem [{\citenamefont {Su}\ \emph {et~al.}(1979)\citenamefont {Su},
  \citenamefont {Schrieffer},\ and\ \citenamefont {Heeger}}]{Su1979}%
  \BibitemOpen
  \bibfield  {author} {\bibinfo {author} {\bibfnamefont {W.~P.}\ \bibnamefont
  {Su}}, \bibinfo {author} {\bibfnamefont {J.~R.}\ \bibnamefont {Schrieffer}},
  \ and\ \bibinfo {author} {\bibfnamefont {A.~J.}\ \bibnamefont {Heeger}},\
  }\href {\doibase 10.1103/PhysRevLett.42.1698} {\bibfield  {journal} {\bibinfo
   {journal} {Phys. Rev. Lett.}\ }\textbf {\bibinfo {volume} {42}},\ \bibinfo
  {pages} {1698} (\bibinfo {year} {1979})}\BibitemShut {NoStop}%
\bibitem [{\citenamefont {Di~Liberto}\ \emph {et~al.}(2017)\citenamefont
  {Di~Liberto}, \citenamefont {Recati}, \citenamefont {Carusotto},\ and\
  \citenamefont {Menotti}}]{Liberto2017}%
  \BibitemOpen
  \bibfield  {author} {\bibinfo {author} {\bibfnamefont {M.}~\bibnamefont
  {Di~Liberto}}, \bibinfo {author} {\bibfnamefont {A.}~\bibnamefont {Recati}},
  \bibinfo {author} {\bibfnamefont {I.}~\bibnamefont {Carusotto}}, \ and\
  \bibinfo {author} {\bibfnamefont {C.}~\bibnamefont {Menotti}},\ }\href
  {\doibase 10.1140/epjst/e2016-60388-y} {\bibfield  {journal} {\bibinfo
  {journal} {Eur. Phys. J. Spec. Top.}\ }\textbf {\bibinfo {volume} {226}},\
  \bibinfo {pages} {2751} (\bibinfo {year} {2017})}\BibitemShut {NoStop}%
\bibitem [{\citenamefont {Qin}\ \emph {et~al.}(2017)\citenamefont {Qin},
  \citenamefont {Mei}, \citenamefont {Ke}, \citenamefont {Zhang},\ and\
  \citenamefont {Lee}}]{Qin2017}%
  \BibitemOpen
  \bibfield  {author} {\bibinfo {author} {\bibfnamefont {X.}~\bibnamefont
  {Qin}}, \bibinfo {author} {\bibfnamefont {F.}~\bibnamefont {Mei}}, \bibinfo
  {author} {\bibfnamefont {Y.}~\bibnamefont {Ke}}, \bibinfo {author}
  {\bibfnamefont {L.}~\bibnamefont {Zhang}}, \ and\ \bibinfo {author}
  {\bibfnamefont {C.}~\bibnamefont {Lee}},\ }\href {\doibase
  10.1103/PhysRevB.96.195134} {\bibfield  {journal} {\bibinfo  {journal} {Phys.
  Rev. B}\ }\textbf {\bibinfo {volume} {96}},\ \bibinfo {pages} {195134}
  (\bibinfo {year} {2017})}\BibitemShut {NoStop}%
\bibitem [{\citenamefont {Qin}\ \emph {et~al.}(2018)\citenamefont {Qin},
  \citenamefont {Mei}, \citenamefont {Ke}, \citenamefont {Zhang},\ and\
  \citenamefont {Lee}}]{Qin2018}%
  \BibitemOpen
  \bibfield  {author} {\bibinfo {author} {\bibfnamefont {X.}~\bibnamefont
  {Qin}}, \bibinfo {author} {\bibfnamefont {F.}~\bibnamefont {Mei}}, \bibinfo
  {author} {\bibfnamefont {Y.}~\bibnamefont {Ke}}, \bibinfo {author}
  {\bibfnamefont {L.}~\bibnamefont {Zhang}}, \ and\ \bibinfo {author}
  {\bibfnamefont {C.}~\bibnamefont {Lee}},\ }\href
  {http://stacks.iop.org/1367-2630/20/i=1/a=013003} {\bibfield  {journal}
  {\bibinfo  {journal} {New J. Phys.}\ }\textbf {\bibinfo {volume} {20}},\
  \bibinfo {pages} {013003} (\bibinfo {year} {2018})}\BibitemShut {NoStop}%
\bibitem [{\citenamefont {Salerno}\ \emph {et~al.}(2018)\citenamefont
  {Salerno}, \citenamefont {Di~Liberto}, \citenamefont {Menotti},\ and\
  \citenamefont {Carusotto}}]{Salerno2018}%
  \BibitemOpen
  \bibfield  {author} {\bibinfo {author} {\bibfnamefont {G.}~\bibnamefont
  {Salerno}}, \bibinfo {author} {\bibfnamefont {M.}~\bibnamefont {Di~Liberto}},
  \bibinfo {author} {\bibfnamefont {C.}~\bibnamefont {Menotti}}, \ and\
  \bibinfo {author} {\bibfnamefont {I.}~\bibnamefont {Carusotto}},\ }\href
  {\doibase 10.1103/PhysRevA.97.013637} {\bibfield  {journal} {\bibinfo
  {journal} {Phys. Rev. A}\ }\textbf {\bibinfo {volume} {97}},\ \bibinfo
  {pages} {013637} (\bibinfo {year} {2018})}\BibitemShut {NoStop}%
\bibitem [{\citenamefont {Corrielli}\ \emph {et~al.}(2013)\citenamefont
  {Corrielli}, \citenamefont {Crespi}, \citenamefont {Della~Valle},
  \citenamefont {Longhi},\ and\ \citenamefont {Osellame}}]{Corrielli2013}%
  \BibitemOpen
  \bibfield  {author} {\bibinfo {author} {\bibfnamefont {G.}~\bibnamefont
  {Corrielli}}, \bibinfo {author} {\bibfnamefont {A.}~\bibnamefont {Crespi}},
  \bibinfo {author} {\bibfnamefont {G.}~\bibnamefont {Della~Valle}}, \bibinfo
  {author} {\bibfnamefont {S.}~\bibnamefont {Longhi}}, \ and\ \bibinfo {author}
  {\bibfnamefont {R.}~\bibnamefont {Osellame}},\ }\href
  {http://dx.doi.org/10.1038/ncomms2578} {\bibfield  {journal} {\bibinfo
  {journal} {Sci. Rep.}\ }\textbf {\bibinfo {volume} {4}},\ \bibinfo {pages}
  {1555} (\bibinfo {year} {2013})}\BibitemShut {NoStop}%
\bibitem [{\citenamefont {Gorlach}\ and\ \citenamefont
  {Poddubny}(2017{\natexlab{b}})}]{Gorlach2017}%
  \BibitemOpen
  \bibfield  {author} {\bibinfo {author} {\bibfnamefont {M.~A.}\ \bibnamefont
  {Gorlach}}\ and\ \bibinfo {author} {\bibfnamefont {A.~N.}\ \bibnamefont
  {Poddubny}},\ }\href {\doibase 10.1103/PhysRevA.95.033831} {\bibfield
  {journal} {\bibinfo  {journal} {Phys. Rev. A}\ }\textbf {\bibinfo {volume}
  {95}},\ \bibinfo {pages} {033831} (\bibinfo {year}
  {2017}{\natexlab{b}})}\BibitemShut {NoStop}%
\bibitem [{\citenamefont {Xiao}\ \emph {et~al.}(2010)\citenamefont {Xiao},
  \citenamefont {Chang},\ and\ \citenamefont {Niu}}]{Xiao2010}%
  \BibitemOpen
  \bibfield  {author} {\bibinfo {author} {\bibfnamefont {D.}~\bibnamefont
  {Xiao}}, \bibinfo {author} {\bibfnamefont {M.-C.}\ \bibnamefont {Chang}}, \
  and\ \bibinfo {author} {\bibfnamefont {Q.}~\bibnamefont {Niu}},\ }\href
  {\doibase 10.1103/RevModPhys.82.1959} {\bibfield  {journal} {\bibinfo
  {journal} {Rev. Mod. Phys.}\ }\textbf {\bibinfo {volume} {82}},\ \bibinfo
  {pages} {1959} (\bibinfo {year} {2010})}\BibitemShut {NoStop}%
\bibitem [{\citenamefont {Liu}\ and\ \citenamefont
  {Wakabayashi}(2017)}]{Liu2017}%
  \BibitemOpen
  \bibfield  {author} {\bibinfo {author} {\bibfnamefont {F.}~\bibnamefont
  {Liu}}\ and\ \bibinfo {author} {\bibfnamefont {K.}~\bibnamefont
  {Wakabayashi}},\ }\href {\doibase 10.1103/PhysRevLett.118.076803} {\bibfield
  {journal} {\bibinfo  {journal} {Phys. Rev. Lett.}\ }\textbf {\bibinfo
  {volume} {118}},\ \bibinfo {pages} {076803} (\bibinfo {year}
  {2017})}\BibitemShut {NoStop}%
\bibitem [{\citenamefont {Delplace}\ \emph {et~al.}(2011)\citenamefont
  {Delplace}, \citenamefont {Ullmo},\ and\ \citenamefont
  {Montambaux}}]{Delplace2011}%
  \BibitemOpen
  \bibfield  {author} {\bibinfo {author} {\bibfnamefont {P.}~\bibnamefont
  {Delplace}}, \bibinfo {author} {\bibfnamefont {D.}~\bibnamefont {Ullmo}}, \
  and\ \bibinfo {author} {\bibfnamefont {G.}~\bibnamefont {Montambaux}},\
  }\href {\doibase 10.1103/PhysRevB.84.195452} {\bibfield  {journal} {\bibinfo
  {journal} {Phys. Rev. B}\ }\textbf {\bibinfo {volume} {84}},\ \bibinfo
  {pages} {195452} (\bibinfo {year} {2011})}\BibitemShut {NoStop}%
\bibitem [{\citenamefont {Hatsugai}(2009)}]{Hatsugai2009}%
  \BibitemOpen
  \bibfield  {author} {\bibinfo {author} {\bibfnamefont {Y.}~\bibnamefont
  {Hatsugai}},\ }\href {\doibase http://dx.doi.org/10.1016/j.ssc.2009.02.055}
  {\bibfield  {journal} {\bibinfo  {journal} {Solid State Commun.}\ }\textbf
  {\bibinfo {volume} {149}},\ \bibinfo {pages} {1061 } (\bibinfo {year}
  {2009})}\BibitemShut {NoStop}%
\bibitem [{\citenamefont {Lau}\ \emph {et~al.}(2015)\citenamefont {Lau},
  \citenamefont {Ortix},\ and\ \citenamefont {van~den Brink}}]{Lau2015}%
  \BibitemOpen
  \bibfield  {author} {\bibinfo {author} {\bibfnamefont {A.}~\bibnamefont
  {Lau}}, \bibinfo {author} {\bibfnamefont {C.}~\bibnamefont {Ortix}}, \ and\
  \bibinfo {author} {\bibfnamefont {J.}~\bibnamefont {van~den Brink}},\ }\href
  {\doibase 10.1103/PhysRevLett.115.216805} {\bibfield  {journal} {\bibinfo
  {journal} {Phys. Rev. Lett.}\ }\textbf {\bibinfo {volume} {115}},\ \bibinfo
  {pages} {216805} (\bibinfo {year} {2015})}\BibitemShut {NoStop}%
\bibitem [{\citenamefont {van Miert}\ \emph {et~al.}(2017)\citenamefont {van
  Miert}, \citenamefont {Ortix},\ and\ \citenamefont {Smith}}]{Miert2017}%
  \BibitemOpen
  \bibfield  {author} {\bibinfo {author} {\bibfnamefont {G.}~\bibnamefont {van
  Miert}}, \bibinfo {author} {\bibfnamefont {C.}~\bibnamefont {Ortix}}, \ and\
  \bibinfo {author} {\bibfnamefont {C.~M.}\ \bibnamefont {Smith}},\ }\href
  {http://stacks.iop.org/2053-1583/4/i=1/a=015023} {\bibfield  {journal}
  {\bibinfo  {journal} {2D Mater.}\ }\textbf {\bibinfo {volume} {4}},\ \bibinfo
  {pages} {015023} (\bibinfo {year} {2017})}\BibitemShut {NoStop}%
\bibitem [{\citenamefont {Kariyado}\ and\ \citenamefont
  {Hatsugai}(2013)}]{Kariyado2013}%
  \BibitemOpen
  \bibfield  {author} {\bibinfo {author} {\bibfnamefont {T.}~\bibnamefont
  {Kariyado}}\ and\ \bibinfo {author} {\bibfnamefont {Y.}~\bibnamefont
  {Hatsugai}},\ }\href {\doibase 10.1103/PhysRevB.88.245126} {\bibfield
  {journal} {\bibinfo  {journal} {Phys. Rev. B}\ }\textbf {\bibinfo {volume}
  {88}},\ \bibinfo {pages} {245126} (\bibinfo {year} {2013})}\BibitemShut
  {NoStop}%
\bibitem [{\citenamefont {Guo}\ \emph {et~al.}(2014)\citenamefont {Guo},
  \citenamefont {Lin},\ and\ \citenamefont {Shen}}]{Guo2014}%
  \BibitemOpen
  \bibfield  {author} {\bibinfo {author} {\bibfnamefont {H.}~\bibnamefont
  {Guo}}, \bibinfo {author} {\bibfnamefont {Y.}~\bibnamefont {Lin}}, \ and\
  \bibinfo {author} {\bibfnamefont {S.-Q.}\ \bibnamefont {Shen}},\ }\href
  {\doibase 10.1103/PhysRevB.90.085413} {\bibfield  {journal} {\bibinfo
  {journal} {Phys. Rev. B}\ }\textbf {\bibinfo {volume} {90}},\ \bibinfo
  {pages} {085413} (\bibinfo {year} {2014})}\BibitemShut {NoStop}%
\bibitem [{\citenamefont {Yoshimura}\ \emph {et~al.}(2014)\citenamefont
  {Yoshimura}, \citenamefont {Imura}, \citenamefont {Fukui},\ and\
  \citenamefont {Hatsugai}}]{Yoshimura2014}%
  \BibitemOpen
  \bibfield  {author} {\bibinfo {author} {\bibfnamefont {Y.}~\bibnamefont
  {Yoshimura}}, \bibinfo {author} {\bibfnamefont {K.-I.}\ \bibnamefont
  {Imura}}, \bibinfo {author} {\bibfnamefont {T.}~\bibnamefont {Fukui}}, \ and\
  \bibinfo {author} {\bibfnamefont {Y.}~\bibnamefont {Hatsugai}},\ }\href
  {\doibase 10.1103/PhysRevB.90.155443} {\bibfield  {journal} {\bibinfo
  {journal} {Phys. Rev. B}\ }\textbf {\bibinfo {volume} {90}},\ \bibinfo
  {pages} {155443} (\bibinfo {year} {2014})}\BibitemShut {NoStop}%
\bibitem [{\citenamefont {Matsumoto}\ \emph {et~al.}(2015)\citenamefont
  {Matsumoto}, \citenamefont {Arita}, \citenamefont {Takane}, \citenamefont
  {Yoshimura},\ and\ \citenamefont {Imura}}]{Matsumoto2015}%
  \BibitemOpen
  \bibfield  {author} {\bibinfo {author} {\bibfnamefont {A.}~\bibnamefont
  {Matsumoto}}, \bibinfo {author} {\bibfnamefont {T.}~\bibnamefont {Arita}},
  \bibinfo {author} {\bibfnamefont {Y.}~\bibnamefont {Takane}}, \bibinfo
  {author} {\bibfnamefont {Y.}~\bibnamefont {Yoshimura}}, \ and\ \bibinfo
  {author} {\bibfnamefont {K.-I.}\ \bibnamefont {Imura}},\ }\href {\doibase
  10.1103/PhysRevB.92.195424} {\bibfield  {journal} {\bibinfo  {journal} {Phys.
  Rev. B}\ }\textbf {\bibinfo {volume} {92}},\ \bibinfo {pages} {195424}
  (\bibinfo {year} {2015})}\BibitemShut {NoStop}%
\bibitem [{\citenamefont {Ryu}\ and\ \citenamefont {Hatsugai}(2002)}]{Ryu2002}%
  \BibitemOpen
  \bibfield  {author} {\bibinfo {author} {\bibfnamefont {S.}~\bibnamefont
  {Ryu}}\ and\ \bibinfo {author} {\bibfnamefont {Y.}~\bibnamefont {Hatsugai}},\
  }\href {\doibase 10.1103/PhysRevLett.89.077002} {\bibfield  {journal}
  {\bibinfo  {journal} {Phys. Rev. Lett.}\ }\textbf {\bibinfo {volume} {89}},\
  \bibinfo {pages} {077002} (\bibinfo {year} {2002})}\BibitemShut {NoStop}%
\bibitem [{\citenamefont {J.~K.~Asb\'oth}\ and\ \citenamefont
  {P\'alyi}(2016)}]{Asboth2016}%
  \BibitemOpen
  \bibfield  {author} {\bibinfo {author} {\bibfnamefont {L.~O.}\ \bibnamefont
  {J.~K.~Asb\'oth}}\ and\ \bibinfo {author} {\bibfnamefont {A.}~\bibnamefont
  {P\'alyi}},\ }\href@noop {} {\emph {\bibinfo {title} {A Short Course on
  Topological Insulators}}}\ (\bibinfo  {publisher} {Springer, Berlin},\
  \bibinfo {year} {2016})\BibitemShut {NoStop}%
\bibitem [{\citenamefont {{Marques}}\ and\ \citenamefont
  {{Dias}}(2017)}]{Marques2017b}%
  \BibitemOpen
  \bibfield  {author} {\bibinfo {author} {\bibfnamefont {A.~M.}\ \bibnamefont
  {{Marques}}}\ and\ \bibinfo {author} {\bibfnamefont {R.~G.}\ \bibnamefont
  {{Dias}}},\ }\href@noop {} {\bibfield  {journal} {\bibinfo  {journal} {ArXiv
  e-prints}\ } (\bibinfo {year} {2017})},\ \Eprint
  {http://arxiv.org/abs/1707.06162} {arXiv:1707.06162 [cond-mat.mes-hall]}
  \BibitemShut {NoStop}%
\bibitem [{\citenamefont {Rossi}(1992)}]{Rossi1992}%
  \BibitemOpen
  \bibfield  {author} {\bibinfo {author} {\bibfnamefont {G.}~\bibnamefont
  {Rossi}},\ }\href {\doibase http://dx.doi.org/10.1016/0379-6779(92)90093-X}
  {\bibfield  {journal} {\bibinfo  {journal} {Synt. Met.}\ }\textbf {\bibinfo
  {volume} {49}},\ \bibinfo {pages} {221 } (\bibinfo {year}
  {1992})}\BibitemShut {NoStop}%
\bibitem [{\citenamefont {Wada}(1992)}]{Wada1992}%
  \BibitemOpen
  \bibfield  {author} {\bibinfo {author} {\bibfnamefont {Y.}~\bibnamefont
  {Wada}},\ }\enquote {\bibinfo {title} {New horizons in low-dimensional
  electron systems: A festschrift in honour of professor h. kamimura},}\ \
  (\bibinfo  {publisher} {Springer Netherlands},\ \bibinfo {address}
  {Dordrecht},\ \bibinfo {year} {1992})\ Chap.\ \bibinfo {chapter} {Doping and
  Disorder in Conducting Polymers}, pp.\ \bibinfo {pages}
  {415--432}\BibitemShut {NoStop}%
\bibitem [{\citenamefont {{Kremer}}\ \emph {et~al.}(2018)\citenamefont
  {{Kremer}}, \citenamefont {{Petrides}}, \citenamefont {{Meyer}},
  \citenamefont {{Heinrich}}, \citenamefont {{Zilberberg}},\ and\ \citenamefont
  {{Szameit}}}]{Kremer2018}%
  \BibitemOpen
  \bibfield  {author} {\bibinfo {author} {\bibfnamefont {M.}~\bibnamefont
  {{Kremer}}}, \bibinfo {author} {\bibfnamefont {I.}~\bibnamefont
  {{Petrides}}}, \bibinfo {author} {\bibfnamefont {E.}~\bibnamefont {{Meyer}}},
  \bibinfo {author} {\bibfnamefont {M.}~\bibnamefont {{Heinrich}}}, \bibinfo
  {author} {\bibfnamefont {O.}~\bibnamefont {{Zilberberg}}}, \ and\ \bibinfo
  {author} {\bibfnamefont {A.}~\bibnamefont {{Szameit}}},\ }\href@noop {}
  {\bibfield  {journal} {\bibinfo  {journal} {ArXiv e-prints}\ } (\bibinfo
  {year} {2018})},\ \Eprint {http://arxiv.org/abs/1805.05209} {arXiv:1805.05209
  [cond-mat.mes-hall]} \BibitemShut {NoStop}%
\bibitem [{Note1()}]{Note1}%
  \BibitemOpen
  \bibinfo {note} {A more in-depth discussion of these results will appear soon
  in a revised version of Ref.~[\protect \rev@citealpnum
  {Marques2017b}]}\BibitemShut {NoStop}%
\bibitem [{Note2()}]{Note2}%
  \BibitemOpen
  \bibinfo {note} {Note that there is a basis rotation that could be performed
  on the diamond chain through which it would be further mapped into an SSH
  chain with dangling sites and renormalized hopping constants, with a similar
  shape to that of Fig.~1(e) in Ref.~[\protect \rev@citealpnum {Xiao2017}].
  However, the problem loses its inversion symmetry in this modified SSH chain
  due to the dangling sites, and therefore our analyses, which relies on the
  existence of inversion symmetry in order to have $\pi $-quantized values for
  the Zak's phases, would not be possible}\BibitemShut {NoStop}%
\bibitem [{\citenamefont {Vargas-Hernández}\ and\ \citenamefont
  {Krems}(2016)}]{Vargas2016}%
  \BibitemOpen
  \bibfield  {author} {\bibinfo {author} {\bibfnamefont {R.~A.}\ \bibnamefont
  {Vargas-Hernández}}\ and\ \bibinfo {author} {\bibfnamefont {R.~V.}\
  \bibnamefont {Krems}},\ }\href
  {http://stacks.iop.org/0953-4075/49/i=23/a=235501} {\bibfield  {journal}
  {\bibinfo  {journal} {J. Phys. B}\ }\textbf {\bibinfo {volume} {49}},\
  \bibinfo {pages} {235501} (\bibinfo {year} {2016})}\BibitemShut {NoStop}%
\bibitem [{\citenamefont {Baier}\ \emph {et~al.}(2016)\citenamefont {Baier},
  \citenamefont {Mark}, \citenamefont {Petter}, \citenamefont {Aikawa},
  \citenamefont {Chomaz}, \citenamefont {Cai}, \citenamefont {Baranov},
  \citenamefont {Zoller},\ and\ \citenamefont {Ferlaino}}]{Baier2016}%
  \BibitemOpen
  \bibfield  {author} {\bibinfo {author} {\bibfnamefont {S.}~\bibnamefont
  {Baier}}, \bibinfo {author} {\bibfnamefont {M.~J.}\ \bibnamefont {Mark}},
  \bibinfo {author} {\bibfnamefont {D.}~\bibnamefont {Petter}}, \bibinfo
  {author} {\bibfnamefont {K.}~\bibnamefont {Aikawa}}, \bibinfo {author}
  {\bibfnamefont {L.}~\bibnamefont {Chomaz}}, \bibinfo {author} {\bibfnamefont
  {Z.}~\bibnamefont {Cai}}, \bibinfo {author} {\bibfnamefont {M.}~\bibnamefont
  {Baranov}}, \bibinfo {author} {\bibfnamefont {P.}~\bibnamefont {Zoller}}, \
  and\ \bibinfo {author} {\bibfnamefont {F.}~\bibnamefont {Ferlaino}},\ }\href
  {\doibase 10.1126/science.aac9812} {\bibfield  {journal} {\bibinfo  {journal}
  {Science}\ }\textbf {\bibinfo {volume} {352}},\ \bibinfo {pages} {201}
  (\bibinfo {year} {2016})}\BibitemShut {NoStop}%
\bibitem [{\citenamefont {Yakubo}\ \emph {et~al.}(2003)\citenamefont {Yakubo},
  \citenamefont {Avishai},\ and\ \citenamefont {Cohen}}]{Yakubo2003}%
  \BibitemOpen
  \bibfield  {author} {\bibinfo {author} {\bibfnamefont {K.}~\bibnamefont
  {Yakubo}}, \bibinfo {author} {\bibfnamefont {Y.}~\bibnamefont {Avishai}}, \
  and\ \bibinfo {author} {\bibfnamefont {D.}~\bibnamefont {Cohen}},\ }\href
  {\doibase 10.1103/PhysRevB.67.125319} {\bibfield  {journal} {\bibinfo
  {journal} {Phys. Rev. B}\ }\textbf {\bibinfo {volume} {67}},\ \bibinfo
  {pages} {125319} (\bibinfo {year} {2003})}\BibitemShut {NoStop}%
\bibitem [{\citenamefont {Wakabayashi}\ and\ \citenamefont
  {Harigaya}(2003)}]{Waka2003}%
  \BibitemOpen
  \bibfield  {author} {\bibinfo {author} {\bibfnamefont {K.}~\bibnamefont
  {Wakabayashi}}\ and\ \bibinfo {author} {\bibfnamefont {K.}~\bibnamefont
  {Harigaya}},\ }\href {https://doi.org/10.1143/JPSJ.72.998} {\bibfield
  {journal} {\bibinfo  {journal} {J. Phys. Soc. Jpn.}\ }\textbf {\bibinfo
  {volume} {72}},\ \bibinfo {pages} {998} (\bibinfo {year} {2003})}\BibitemShut
  {NoStop}%
\end{thebibliography}%
\cleardoublepage
\appendix

\section{Fundamental domain of the mapped model for spinless fermions}
\label{sec:appendix}

\begin{figure*}[h]
	\begin{center}
		\includegraphics[width=0.95 \textwidth, height=0.30 \textheight]{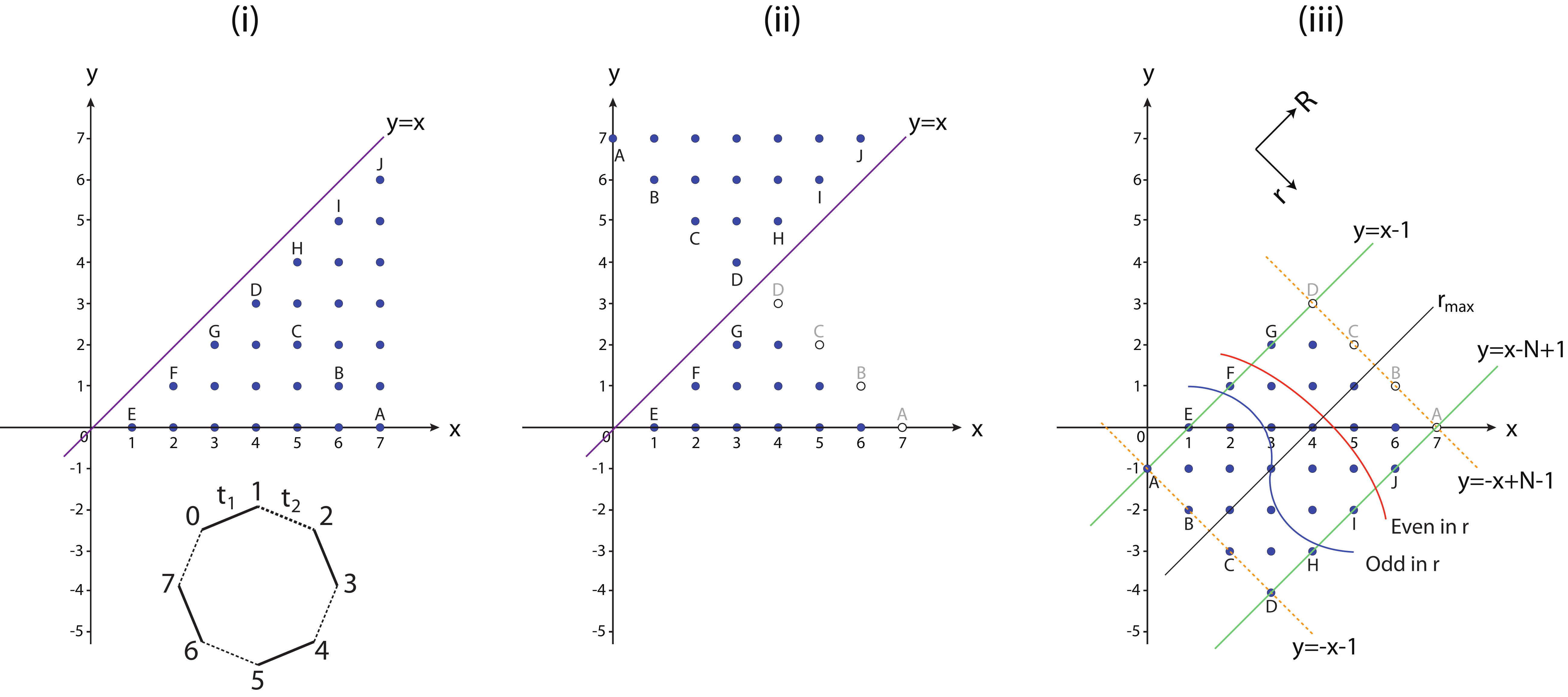}
	\end{center}
	\caption{(i) Blue dots correspond to the available states, in the mapped model in $xOy$, of a system of two spinless fermions in an SSH ring with N=8 sites, shown at the bottom.
	(ii) States on the DA diagonal and above are projected onto their anti-symmetric counterparts.
(iii) Taking advantage of the periodicity of the states, the $y$ coordinate of the projected states of (ii) is shifted by $-N$.
Green solid lines represent the open boundaries in the $r$ direction and equally labeled sites at the orange dashed lines are the same site in the periodic $R$ direction.
Red and blue curves represent possible even and odd solutions along $r$, respectively, for a wavefunction $\psi(R,r)$ in this fundamental domain.}
	\label{fig:funddom}
\end{figure*}

We present here the procedure to established the fundamental domain for the case of two non-interacting spinless fermions in an SSH rings with $q=N/2$ even. 
We consider the SSH ring with $N=8$ sites of Fig.~\ref{fig:2dmapping}(c).
One starts by establishing a numbering of the sites of the original model, which in our choice goes from 0 to $N-1$ [see bottom of Fig.~\ref{fig:funddom}(i)].
By assigning the position of each particle to a different degree of freedom (directions $x$ and $y$, with $x>y$ ordering, that is, $y=0,...,N-2$ and $x=y+1,y+2,...,N-1$), such that $\ket{x,y}$ specifies a given two-particle state in the original model, one can identify all possible states of this model in the $xOy$ plane, as in Fig.~\ref{fig:funddom}(i). 
States along the $y=x$ line, representing same site occupation, are forbidden due to Pauli's exclusion principle (they should be considered in the bosonic and distinguishable particles cases).
Since we are only interested here in counting the number of states, we omit the alternating hopping constants between states in the $xOy$ plane, which would fully map the problem into the 2D SSH model of Fig.~\ref{fig:2dmapping}(b).

Having counted all available states, we take advantage of the anti-symmetry of the states with respect to $y=x$ to project the states on the DA diagonal and above onto their anti-symmetric counterparts, as exemplified in Fig.~\ref{fig:funddom}(ii).
Recalling that the states have a periodicity defined by $\ket{x,y}\equiv\ket{x\pm iN,y\pm jN}$, with $(i,j)\in \mathbb{Z}\times\mathbb{Z}$, we shift the projected states by $(i,j)=(0,-1)$ and get the final form of the mapped 2D model shown in Fig.~\ref{fig:funddom}(iii), where a wavefunction $\psi(R,r)$ satisfies open boundary conditions in $r$ and M\"obius boundary conditions \cite{Yakubo2003,Waka2003} (MBC) in $R$ (note the inverted A-B-C-D sequence at the orange dashed lines of the periodic $R$ direction),
\begin{eqnarray}
\psi(R,0)=\psi(R,N)=0,
\\
\psi(R,r)=\psi(R+N/2,N-r),
\label{eq:mobiusbc}
\end{eqnarray}
with $R=\frac{x+y}{2}$ varying in half-integer steps.  
When $\psi(R,r)$ is even along $r$ [see red curve in Fig.~\ref{fig:funddom}(iii)], cylinder boundary conditions (CBC) and MBC become equivalent and (\ref{eq:mobiusbc}) can therefore rewritten as
\begin{equation}
\psi(R,r)=\psi(R+N/2,r).
\end{equation} 
When $\psi(R,r)$ is odd along $r$ [see blue curve in Fig.~\ref{fig:funddom}(iii)], on the other hand, one has to introduce a $\pi$ phase shift in the boundary condition to recover CBC \cite{Yakubo2003,Waka2003},
\begin{equation}
\psi(R,r)=-\psi(R+N/2,r),
\end{equation}
which creates a $\pi$ magnetic flux along $r$ that induces a shift in the center-of-mass momentum $K$ proportional to the system size $N$.
The general relation,  when one switches from MBC to CBC, can be encapsulated as 
\begin{equation}
\label{eq:kshift}
\begin{cases}
K\to K, \ \ \ \ \ \ \ \ \ \ \ \mbox{for $\psi$ even in $r$},
\\
K\to K+\frac{2\pi}{N}, \ \ \ \ \mbox{for $\psi$ odd in $r$}.
\end{cases}
\end{equation}
For large $N$, the $K$ (energy) shift which affects only states with odd $k$ becomes negligible, and one can consider only the even solution in (\ref{eq:kshift}) for all $k$.

The procedure to determine the fundamental domain in the bosonic case is the same as for the case of spinless fermions shown here, except that one also has to consider the sites along the $y=x$ ($r=0$) line and an extra $\sqrt{2}$ factor at the hopping terms connected to the sites at both edges in $r$, when one considers the states of the fundamental domain to be symmetrized, as explained in the discussion around (\ref{eq:symbosons}) in the main text.
To determine the fundamental domain for the case of two distinguishable particles, one only has to identify all possible states and impose periodic boundary conditions in both the $r$ and $R$ directions (which produces a toroidal shape).
The unit cell of this case is shown in the light green region in Fig.~\ref{fig:2dmapping}(b).
The full torus can then be constructed by adding a second adjacent unit cell and wrapping also around $R$.

\begin{figure*}[h]
	\begin{center}
		\includegraphics[width=0.75 \textwidth, height=0.25 \textheight]{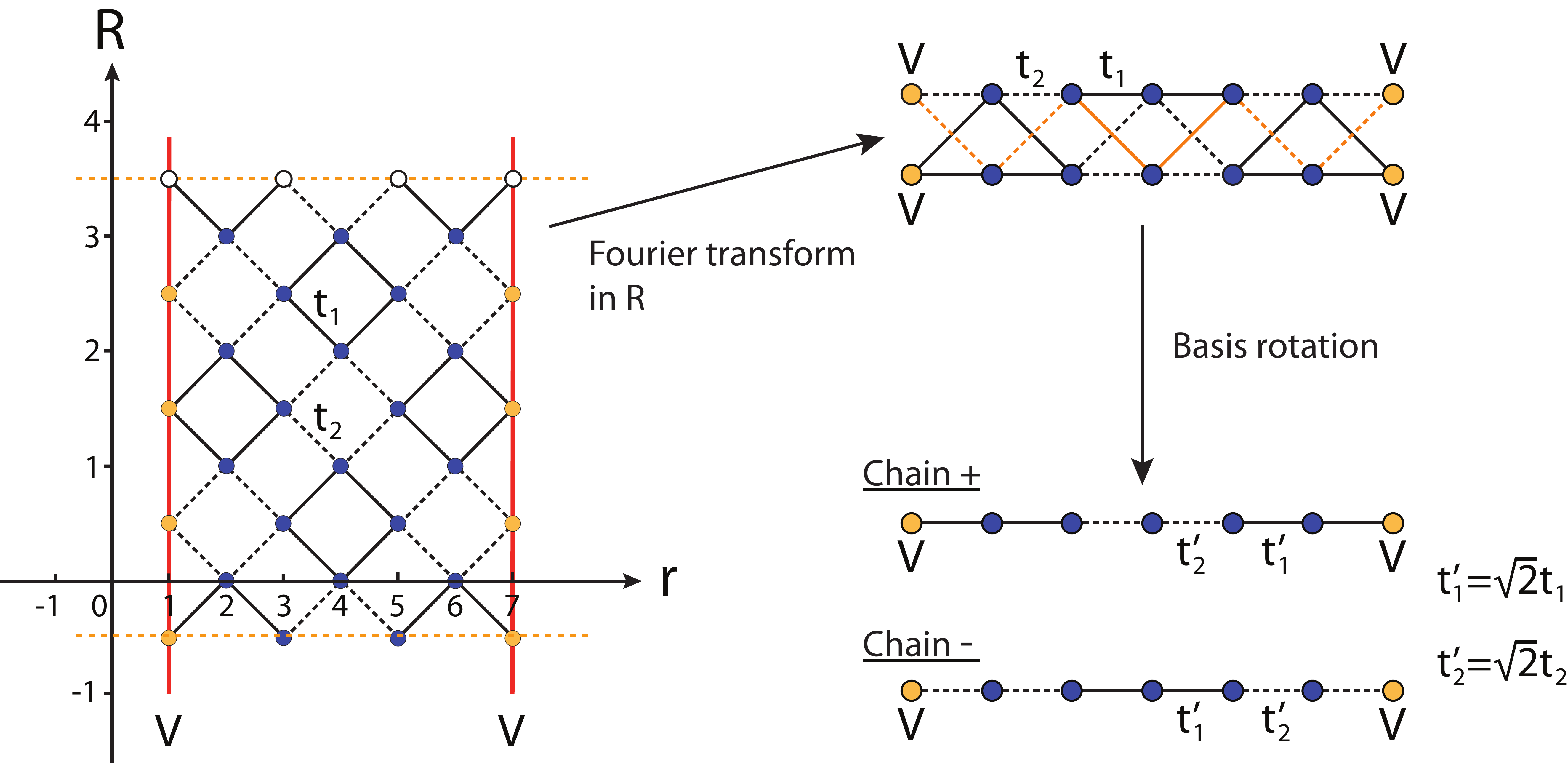}
	\end{center}
	\caption{At the left, we consider the fundamental domain of Fig.~\ref{fig:funddom}(iii) with the inclusion of an on-site potential $V$ at the edge sites in $r$. 
			By Fourier transforming this fundamental domain in the periodic $R$-direction using CBC and considering then $K=\pi$, we arrive at the two-leg ladder on the right top. 
			Orange couplings carry an extra $\pi$ phase.
			Under the basis rotation of (\ref{eq:basisrot1}-\ref{eq:basisrot2}), the two-leg ladder is transformed into the two decoupled chains with renormalized couplings labeled ``chain +'' and ``chain -''.}
	\label{fig:fdtot1t1t2t2}
\end{figure*}

For the sake of completeness, we take the fundamental domain of Fig.~\ref{fig:funddom}(iii), considering now an on-site potential $V$ at the edge sites in $r$ and CBC, and complete in Fig.~\ref{fig:fdtot1t1t2t2} the steps that transform it into the $t_1t_1t_2t_2$ model for $K=\pi$.
Alternatively, Fig.~\ref{fig:fdtot1t1t2t2} can also be seen as the continuation of Fig.~\ref{fig:2dmapping}(c).

\end{document}